\documentclass[sigconf]{acmart}

\AtBeginDocument{%
  \providecommand\BibTeX{{%
    \normalfont B\kern-0.5em{\scshape i\kern-0.25em b}\kern-0.8em\TeX}}}

\copyrightyear{2026}
\acmYear{2026}
\setcopyright{cc}
\setcctype{by}
\acmConference[CHI '26]{Proceedings of the 2026 CHI Conference on Human Factors in Computing Systems}{April 13--17, 2026}{Barcelona, Spain}
\acmBooktitle{Proceedings of the 2026 CHI Conference on Human Factors in Computing Systems (CHI '26), April 13--17, 2026, Barcelona, Spain}
\acmPrice{}
\acmDOI{10.1145/3772318.3790618}
\acmISBN{979-8-4007-2278-3/2026/04}

\usepackage{physics}
\usepackage{array}
\usepackage{enumitem}
\usepackage{booktabs}

\usepackage{graphicx}
\usepackage{amsmath} 
\usepackage{xfrac}
\usepackage{subcaption} 

\usepackage{xcolor}

\definecolor{myblue}{rgb}{0.345, 0.604, 0.839}
\definecolor{mypurple}{rgb}{0.749, 0.341, 0.733}

\newcommand{\new}[1]{\textcolor{black}{#1}}

\newcommand{\rev}[1]{\textcolor{black}{#1}}

\newcommand{\removegap}[0]{\vspace{-0.2cm}}
\newcommand{\nonacmgap}[0]{\vspace{-0.15cm}}
\newcommand*\dsl[1]{\textcolor{black}{\small{\texttt{#1}}}}

\begin{document}

\title[Gesturing Toward Abstraction]{Gesturing Toward Abstraction: Multimodal Convention Formation in Collaborative Physical Tasks}

\author{Kiyosu Maeda}
\orcid{0000-0002-3270-1974}
\email{km9567@princeton.edu}
\affiliation{%
  \institution{Princeton University}
  \city{Princeton}
  \state{NJ}
  \country{USA}
}

\author{William P. McCarthy}
\orcid{0000-0002-3084-7152}
\email{wmccarthy@ucsd.edu}
\affiliation{%
  \institution{UC San Diego}
  \city{La Jolla}
  \state{CA}
  \country{USA}
}

\author{Ching-Yi Tsai}
\orcid{0000-0001-5664-6562}
\email{ching-yi@princeton.edu}
\affiliation{%
  \institution{Princeton University}
  \city{Princeton}
  \state{NJ}
  \country{USA}
}

\author{Jeffrey Mu}
\orcid{0009-0009-8925-8596}
\email{jeffrey_mu@brown.edu}
\affiliation{%
  \institution{Brown University}
  \city{Providence}
  \state{RI}
  \country{USA}
}

\author{Haoliang Wang}
\orcid{0000-0001-7529-6991}
\email{hlwang@mit.edu}
\affiliation{%
  \institution{MIT}
  \city{Cambridge}
  \state{MA}
  \country{USA}
}

\author{Robert D. Hawkins}
\orcid{0000-0001-9089-8544}
\email{rdhawkins@stanford.edu}
\affiliation{%
  \institution{Stanford University}
  \city{Stanford}
  \state{CA}
  \country{USA}
}

\author{Judith E. Fan}
\orcid{0000-0002-0097-3254}
\email{jefan@stanford.edu}
\affiliation{%
  \institution{Stanford University}
  \city{Stanford}
  \state{CA}
  \country{USA}
}

\author{Parastoo Abtahi}
\orcid{0009-0000-2145-3445}
\email{parastoo@princeton.edu}
\affiliation{%
  \institution{Princeton University}
  \city{Princeton}
  \state{NJ}
  \country{USA}
}

\renewcommand{\shortauthors}{Maeda et al.}

\settopmatter{authorsperrow=4}

\begin{abstract}
    A quintessential feature of human intelligence is the ability to create ad hoc conventions over time to achieve shared goals efficiently. We investigate how communication strategies evolve through repeated collaboration as people coordinate on shared procedural abstractions. To this end, we conducted an online unimodal study (n = 98) using natural language to probe abstraction hierarchies. In a follow-up lab study (n = 40), we examined how multimodal communication (speech and gestures) changed during physical collaboration. Pairs used augmented reality to isolate their partner’s hand and voice; one participant viewed a 3D virtual tower and sent instructions to the other, who built the physical tower. Participants became faster and more accurate by establishing linguistic and gestural abstractions and using cross-modal redundancy to emphasize key changes from previous interactions. Based on these findings, we extend probabilistic models of convention formation to multimodal settings, capturing shifts in modality preferences. Our findings and model provide building blocks for designing convention-aware intelligent agents situated in the physical world.
\end{abstract}


    


\begin{teaserfigure}
    \includegraphics[width=\textwidth]{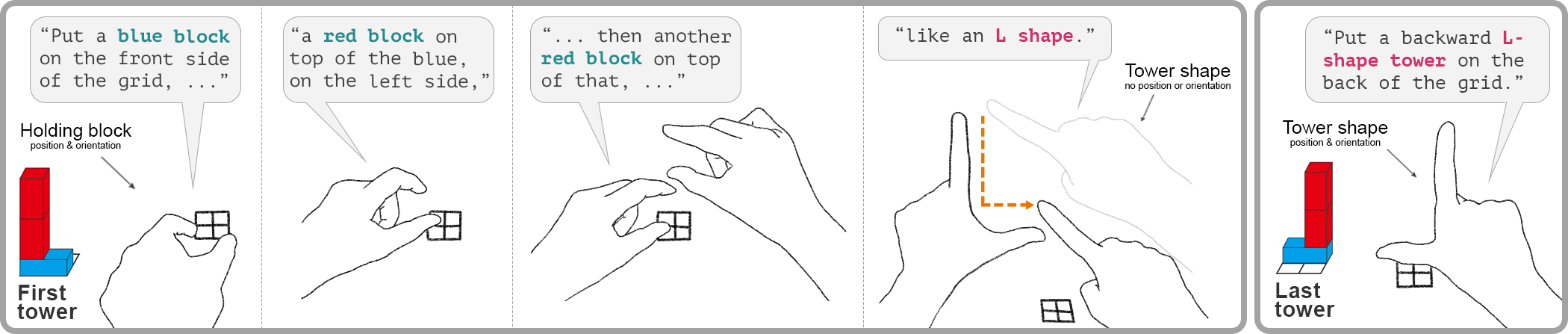}
    \caption[From block-level to abstract instructions]{We study how people instruct a partner in a physical assembly task using speech and gesture. Participants begin with block-level instructions (low abstraction), then establish linguistic and gestural conventions and use them to describe the overall tower shape (high abstraction), while employing cross-modal redundancy to emphasize changes in position and orientation. \looseness=-1}
    \Description{Five-panel teaser showing a shift from block-by-block to abstract tower descriptions. Panel 1 shows the first L-shaped tower made from three LEGO blocks (blue base, two red blocks stacked). A speech bubble says Put a blue block on the front side of the grid, with a hand precisely placing an imaginary block on a two-by-two grid. Panel 2 shows a speech bubble saying a red block on top of the blue, on the left side, with a hand holding an imaginary block vertically above the previous position. Panel 3 shows a speech bubble saying then another red block on top of that, with the right hand stacking another imaginary block. Panel 4 shows a speech bubble saying like an L shape. Two hands depict an L-shape gesture representing tower shape without position or orientation. The final panel shows the same tower in a different position and orientation, with a speech bubble reading Put a backward L-shape tower on the back of the grid and a hand indicating the back row of the grid.}
    \vspace{0.4cm}
    \label{fig:teaser}
\end{teaserfigure}
    
\maketitle

\section{Introduction}
    
    

    People leverage multimodal signals to convey their intentions and understand others' intentions. While people can naturally use these cues in human-human communication, supporting multimodal \new{communication} with agents (e.g., \new{AI-powered wearables} and robots) is a challenging problem.
    Prior works have proposed AI systems that interpret multimodal cues\rev{~\cite{LeeCHI24, pustejovsky2017}} and communicate their intentions in predictable and legible ways during one-time interactions~\cite{endeHPWHA11}. However, across repeated interactions, people form ad hoc conventions with their collaborators, using more concise expressions to refer to objects or concepts~\cite{krauss1964changes} and convey more abstract or chunked information to others. While \rev{these} signals become increasingly ambiguous, sacrificing some informativeness~\cite{wilkes1992coordinating}, partners still communicate accurately and efficiently based on \rev{their shared} history.\looseness=-1

    To date, most controlled studies have been focused on unimodal scenarios~\cite{fay2010interactive, garrod1994conversation, hawkins2023visual} or 2D collaborative tasks~\cite{jayannavarNH20, mccarthy2024communicating}. 
    Yet, many real-world tasks require multiple agents to coordinate their behavior over time in physical settings, using their hands to convey information that is not expressed in speech.
    For example, people frequently say ambiguous words (e.g., ``this'') while referring to an object with a deictic gesture~\cite{bolt80}. Hands also communicate the shape of an object, describe the distance, and demonstrate actions~\cite{bekkerOO95, fussellS0OMK04}. Understanding how such multimodal signals \textit{evolve} during iterated collaboration in physical settings is crucial for designing agents that can form and use multimodal conventions.
    
    Our study examined how people use voice and gestures to establish conventions and adapt multimodal communication over repeated physical collaboration.
    \emph{First}, we ran an online \emph{unimodal} study to probe how partners coordinate on linguistic abstractions; dyads shifted from block-based descriptions to concise tower-level expressions, reducing instruction length over repetitions. \emph{Then}, we conducted a controlled AR lab study that isolated speech and gestures while participants built physical towers. Results indicated that participants' instructions were shortened by \new{forming} gestural and linguistic abstractions \new{and} shifting from block-based to tower-based instructions\new{.}
    We also found that shared abstractions were introduced near the end of the first \new{exchange} and later used at the beginning of instructions. Furthermore, redundancy in multimodal signals increased for tower-based instructions; participants used both modalities to describe changes in the position and orientation of target towers, \rev{emphasizing} variations across repetitions.

    Based on our observations, we built multimodal computational agents that probabilistically reason about each other's beliefs and continuously learn abstract \new{tower} representations in our task setting. We extended the Rational Speech Act (RSA) framework~\cite{frank2012predicting, goodman2016pragmatic} to develop a model of convention formation~\cite{hawkins2023partners} that can operate over a multimodal lexicon (i.e., a mapping between symbols and utterances/gestures) and consider different combinations of multimodal signals (i.e., redundant, complementary, and language-only) by assuming that the lexicon returns continuous real values~\cite{degen2020redundancy}. 
    
    \rev{We conducted simulations to answer two research questions: \textbf{(RQ1)} Can agents acquire abstract tower representations over repeated interaction? and \textbf{(RQ2)} can the model capture how modality use shifts for different participants?} The model successfully acquired abstract programs and captured modality-dependent behavioral shifts across repetitions. The proposed computational models and the findings from the study will pave the way to design multimodal agents that form conventions with humans for efficient and successful iterated collaboration in physical tasks. \looseness=-1

\noindent \textbf{Contributions:} \nonacmgap
    \begin{itemize}
        \item Large-scale \textbf{online unimodal study} investigating natural language communication over repeated interactions.
        \item AR-mediated \textbf{multimodal lab study} examining gesture and speech communication, including a multimodal dataset to support future research \rev{in repeated physical collaboration}. 
        \item \textbf{Computational model} and simulation, exhibiting behaviors aligned with study findings, with a flexible design extendable to more synchronous and symmetric communication. \looseness=-1
        \nonacmgap
    \end{itemize}

\section{Related Work}    

\subsection{Gestures in Thought and Communication}
    \rev{Gestures play a critical role in cognitive processes when thinking and speaking} and have implications for both learning and communication~\cite{kita2017gestures, mcNeill_1992, goldin2005hearing, kessell2006using, kang2016hands}.
    Gestures \rev{have been shown to enhance spatial reasoning ~\cite{chu2011nature, goldin2005our} and} improve problem solving and concept learning~\cite{beilock2010gesture}. \rev{They can promote knowledge transfer~\cite{novack2014action} and convey visual concepts difficult to express in speech ~\cite{alibali1999illuminating}.}
    Gestures \rev{also} facilitate language comprehension by adding information \rev{to} speech. People \rev{use} \textit{deictic} gestures to refer to objects or concepts~\cite{chen0KKZGZH21, jiangXXLPZZ23}, \textit{symbolic} gestures \rev{for} abstract meanings~\cite{makriniMLV17}, and \textit{iconic} gestures to disambiguate utterances~\cite{holle2007role}.

    \rev{Gestures change over extended interaction.} In mental rotation tasks, people initially use gestures that simulate direct manipulation of objects \rev{and} later \rev{represent abstract object motions}~\cite{chu2008spontaneous}\rev{, and gesture frequency also changes over time with changes in strategy}~\cite{schwartz1996shuttling}. In iterated communication, gestures become more efficient and less kinematically complex~\cite{pouw2021systematic}. 
    Between partners, repeated interaction leads to convergence on shared gestural conventions~\cite{fay2014creating}. People adapt to \rev{each other} \rev{and} synchronizing their motion through \textit{gesture entrainment} ~\cite{anserminMBG16,kimotoISTSH16,ono2001model,stoeva2024body}\rev{, and systematic structures can emerge over time within a community}~\cite{sandler2005emergence}. 

    However, gestures rarely occur in isolation and typically co-occur with speech, distributing information across modalities\rev{~\cite{goldin2006talking,mcNeill_1992,quek2002multimodal}}. We investigate \rev{how people use} speech and gesture \rev{to communicate} during physical assembly~\cite{zheng2024putting}, examining how information is allocated across channels \rev{and} how it changes over time.\looseness=-1

\subsection{Multimodal Cues in Collaborative Tasks}
    Multimodal signals are fundamental to human communication. As computing systems become more integrated into physical environments, it is crucial to interpret human intent from multimodal cues\rev{~\cite{bolt80, pustejovsky2017}} and to generate information, not only through text but also with embodied output~\cite{sharmaPH98}. Extensive research has focused on enabling agents to interpret and present multimodal information, thereby enhancing human-computer communication. On the input side, understanding co-speech gestures has been used to specify spatiotemporal references in VR~\cite{hu2025gesprompt}, disambiguate queries in AR~\cite{LeeCHI24}, generate live visual effects~\cite{saquibKWL19}, and provide robot assistance~\cite{LinCHXS23, robinson2023robotic}. On the output side, systems have presented multimodal robot intent~\cite{gielniakT11} and embodied instructions for navigation~\cite{liu2001comparative}, music learning~\cite{karolusS0W23}, and cooking~\cite{yangQCSBLL24}. \looseness=-1         

    Analyzing human-human communication \rev{provides insights} for these applications~\cite{endeHPWHA11, fussellKS00}. 
    Narayana et al.~\cite{narayana2018cooperating} analyzed gesture use in collaborative tasks, identifying common gesture types and how their use varies with speech availability.
    Gergle et al.~\cite{gergleKF04, gergle2004language} investigated how visual information affects \rev{collaboration} in 2D puzzle tasks, building on the conversational grounding framework~\cite{clark1996using} and the principle of least collaborative effort~\cite{clark1986referring}. \rev{They found that with shared visual context,} collaborators use visual cues (e.g., pointing, moving pieces) \rev{rather than} explicit verbal \rev{grounding}.
    \rev{In assembly tasks, multimodal communication significantly reduced completion time compared to speech alone~\cite{wang2021sa, wang2017exploring}.} 
    Gleeson et al.~\cite{gleesonMHCA13} \rev{analyzed these interactions to derive a} gesture \rev{vocabulary} and develop a robotic arm that communicates through gestures. 
    
    Recent studies have captured datasets of human–human communication (e.g., EGGNOG~\cite{wangFNPM17}, HoloAssist~\cite{wangKRPCABFTFJP23}). \rev{We build on these by conducting a more controlled experiment that carefully isolates speech and gesture using AR and limits feedback between collaborators; this is needed for analysis of information across modalities and to support computational modeling.} \rev{Moreover, representations have been introduced to} encode gesture semantics and support multimodal understanding~\cite{brutti2022abstract}. \rev{Prior simulations have also interpreted gesture and language instructions using predefined symbolic mappings to update beliefs about the physical world~\cite{pustejovsky2017, krishnaswamy2017communicating, krishnaswamy2022voxworld, krishnaswamy2020diana}.} 
    \rev{However, there is limited work on modeling or simulating multimodal human–human communication in \textit{repeated interactions} where collaborators adapt and coordinate their behaviors over time.}
    We study how multimodal communication signals change during repeated collaborative physical tasks \rev{and model how beliefs about semantic conventions are updated.}

\subsection{Conventions in Repeated Collaboration}
    In repeated interactions, people form ad hoc conventions---solutions to recurring coordination problems with others~\cite{lewis2008convention}. For instance, people often shorten referring expressions to reduce the effort required to identify objects~\cite{boyce2023communicative, stewartYE20} or align their word choices with those of their conversational partner when referring to the same object repeatedly, known as \textit{lexical entrainment}~\cite{brennan1996conceptual, iioSSMAH09, phukon2023effect}. People also chunk complicated steps into a single instruction to share their goals efficiently~\cite{hawkinsKSG20}. 
    While those in different groups may struggle to understand others' conventions due to ambiguity~\cite{wilkes1992coordinating}, members within a group can readily disambiguate intentions by updating their expectations based on past interactions~\cite{clark1986referring}. \rev{Thus, conventions are essential for successful and efficient collaboration.}\looseness=-1
    

    Prior work in cognitive science has explored how conventions form~\cite{fried2022pragmatics, garrod1994conversation, wangLM16} \rev{across a range of collaborative tasks, including} reference games~\cite{boyce2024}. \rev{Over repeated interactions, people develop increasingly concise linguistic expressions~\cite{hawkins2023partners, krauss1964changes} and graphical depictions~\cite{fay2010interactive, hawkins2023visual, mccarthy2024communicating}, enabling more efficient communication.}
    \rev{This understanding has motivated computational approaches that model how AI agents can establish conventions~\cite{lazaridou-abs-2006-02419, qiuXFGJZZ22}.} For example, Hua et al.~\cite{hua2024talk} show that current vision–language models rarely form conventions with users and propose methods to support such linguistic adaptation~\cite{hua2025posttraining}. Shih et al.~\cite{shihSKES21} separate convention-dependent and rule-based behaviors during iterated collaboration to acquire conventions. However, most of these efforts focus on unimodal, text-based communication. There is limited work on multimodal conventions in multi-step physical tasks. \looseness=-1

    To address this, we first establish our experimental paradigm in a unimodal online study~\cite{mccarthy2021learning} and then conduct a lab study to investigate multimodal communication during repeated physical assembly~\cite{maeda2025using}. Our findings and computational model aim to explain how people form multimodal abstractions when giving instructions on physical tasks and how they adjust the amounts of information across modalities (speech and gesture) over time.\looseness=-1 
    


    

        

\section{Online Unimodal Study}

The goal of our first, \rev{online unimodal} study was to establish a paradigm for investigating how people simultaneously coordinate on a shared set of concepts \rev{and on} a way of communicating about them.
We explored this phenomenon in an assembly domain \cite{bapst2019structured, mccarthy2020blocks,walsman2022break,bramley2023active} where participants encountered visual scenes populated by a recurring set of block towers. 
The scenes were hierarchically organized and could be represented at multiple levels of abstraction\rev{---for instance, either as whole structures or as combinations of simpler units}. 
As participants \rev{viewed} multiple scenes, we hypothesized that certain ``chunks'' would be preferred, grouping primitive elements (individual blocks) into more complex configurations~ \cite{aslin1998computation,christiansen2016now,austerweil2013nonparametric}.
However, these newly formed abstractions are only useful in this task if they can be successfully communicated to others, which requires overcoming the inherent risk of miscommunication that accompanies the use of new linguistic terms.\looseness=-1

\subsection{Method}
\subsubsection{Participants}
We recruited $146$ participants from Amazon Mechanical Turk and paired them into 73 dyads \rev{for} our IRB-approved study.
We excluded 24 dyads who failed to meet preregistered criteria ($\geq75$\% reconstruction accuracy on $\geq75$\% of trials, \rev{or} self-reported confusion \rev{or} non-fluency in English).
The session lasted 30–50 minutes. Participants \rev{received} a minimum compensation of \$5.00 plus a performance bonus of up to \$3.00. 


\begin{figure}[h!]
    \centering
    \includegraphics[width=\linewidth]{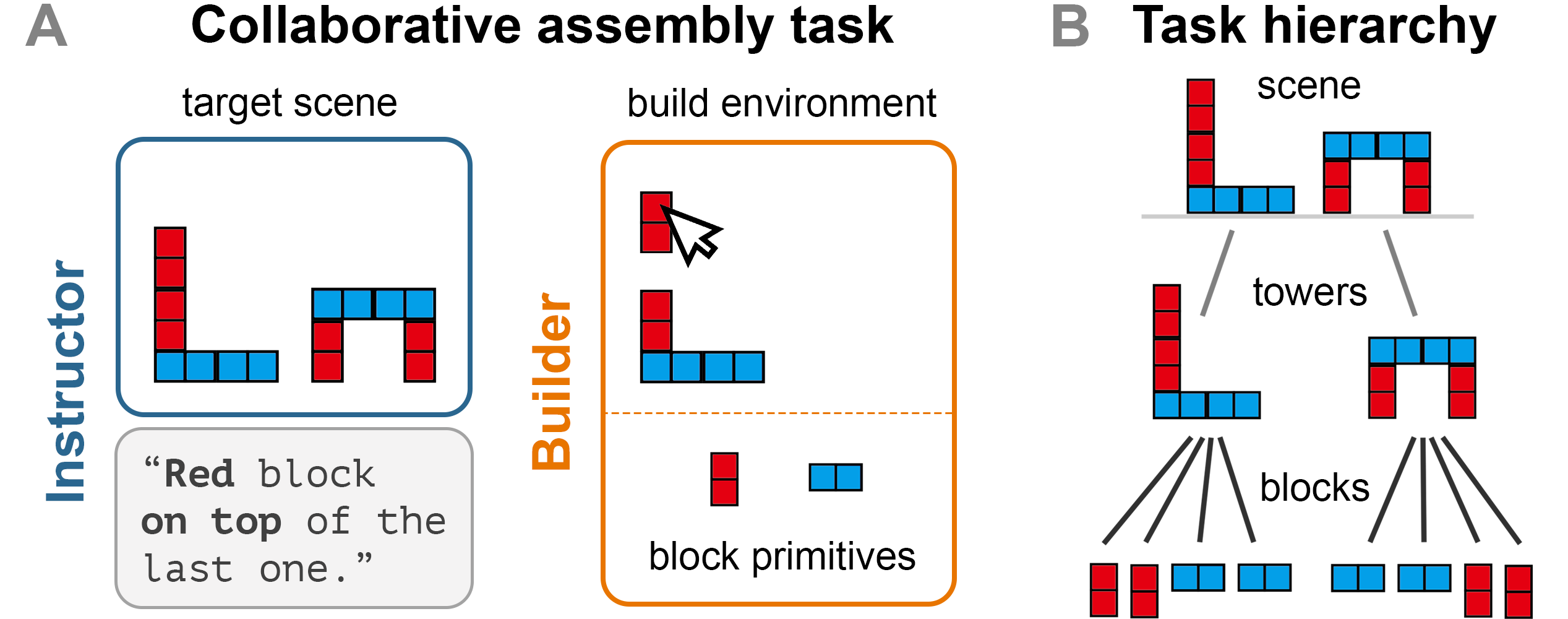}
    \caption[Study 1 task setup]{(A) Instructor viewed a target scene and gave assembly instructions to the Builder. (B) Scenes with two towers.}
    \Description{Panel A shows a collaborative assembly task. The Instructor views a target scene containing an L-shaped tower and an upside-down U-shaped tower made from red and blue blocks, with a speech bubble saying Red block on top of the last one. The Builder panel shows the build environment with block primitives, including a two-by-one vertical red block and a one-by-two horizontal blue block. Panel B shows a task hierarchy diagram with a scene at the top, two towers in the middle, and eight domino-shaped blocks at the bottom, illustrating that each tower is composed of four blocks.}
    \label{fig:study1_setup}
\end{figure}

\begin{figure*}[!t]
    \centering
    \includegraphics[width=\textwidth]{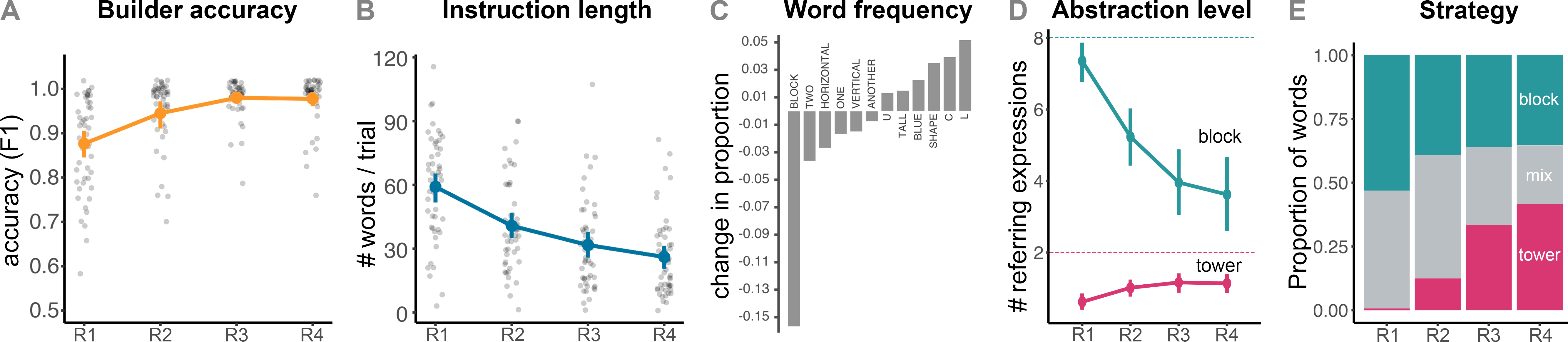}
    \caption[Study 1 results]{(A) Mean reconstruction accuracy improved across repetitions. (B) Mean instruction length per trial decreased across repetitions as dyads became more effective at collaborating. (C) Words with the largest positive or negative changes in frequency from R1 to R4. (D) Change in the number of block- and tower-level references. Dashed lines are the maximum possible number of blocks and towers. (E) The proportion of expressions exclusively referring to blocks or towers. Error bars: 95\% CIs.}
    \Description{Results across repetitions R1-R4 in five panels. A) Builder accuracy (F1) increases from about 0.90 to about 0.98. B) Instruction length (number of words per trial) decreases from about 60 to about 30. C) Word frequency change in proportion shows decreases for BLOCK, TWO, HORIZONTAL, VERTICAL, ANOTHER, and increases for L, C, SHAPE, BLUE. D) Abstraction level (number of referring expressions) shows block references decreasing and tower references increasing slightly. E) Strategy stacked bars show the proportion of words shifting from block to more towers by R4.}
    \label{fig:study1_all_results}
\end{figure*}

\subsubsection{Stimuli}
Each scene consisted of two towers built hierarchically from four domino-shaped blocks---two vertical and two horizontal (\autoref{fig:study1_setup}B).
We created three unique towers and used a \emph{repeated} design where each tower appeared multiple times.
All three tower pairs appeared in randomized order across four \emph{repetitions}, \rev{yielding} twelve trials. Each tower appeared equally on the left and right, ensuring no association between towers and their position.\looseness=-1   

Each participant was assigned a fixed role of \textit{Instructor} or \textit{Builder} and \rev{completed} twelve trials.
In each trial, the Instructor \rev{saw} a target scene \rev{with} block towers (\autoref{fig:study1_setup}A), \rev{while the Builder viewed an empty grid for placing} domino-like blocks.
The Instructor \rev{provided} step-by-step assembly instructions through a free-response text box, \rev{and} the Builder used \rev{these to} reconstruct the scene.
\rev{They took turns as needed: on each Instructor turn, they sent one message (up to 100 characters), and on each Builder turn, they placed any number  ($\geq0$) of blocks before selecting ``done.''} 
\rev{Blocks had to be supported from below and could not be moved once placed.}
The Instructor \rev{saw the Builder's block placements} in real time, but communication was otherwise unidirectional. 
\rev{A 30-second countdown appeared on each turn to encourage quick progress; exceeding the limit had no penalty.}
After all eight blocks were placed, participants received feedback on the mismatch between the target and reconstructed scenes before advancing to the next trial.



\subsection{Results}


\subsubsection{Reconstruction accuracy improves across repetitions}
Although each interaction spanned only twelve trials, \rev{\textbf{(H1)}} we hypothesized that dyads would rapidly develop shared task representations, \rev{leading to more successful} collaboration. 
We first \rev{verified} that dyads \rev{could} perform the assembly task.
We measured performance using reconstruction accuracy, quantified as the $F_1$ overlap between the reconstructed tower and the target silhouette, which captures both missing blocks (recall) and extraneous ones (precision).
$F_1$ ranges from 0 (no overlap) to 1 (perfect). Initial reconstructions were accurate (mean $F_1=0.88$, 95\% CI $=[0.85,0.90]$), roughly corresponding to one misplaced block, and final reconstructions were near ceiling ($F_1=0.98$, 95\% CI $=[0.96,0.99]$).     
A linear mixed-effects model predicting $F_1$ from repetition number, with random intercepts and slopes by dyad, revealed a significant improvement across repetitions ($\beta=0.92$, $t(54.84)=6.22$, $p< .001$; \autoref{fig:study1_all_results}A).

\begin{figure}[b!]
    \centering
    \includegraphics[width=\linewidth]{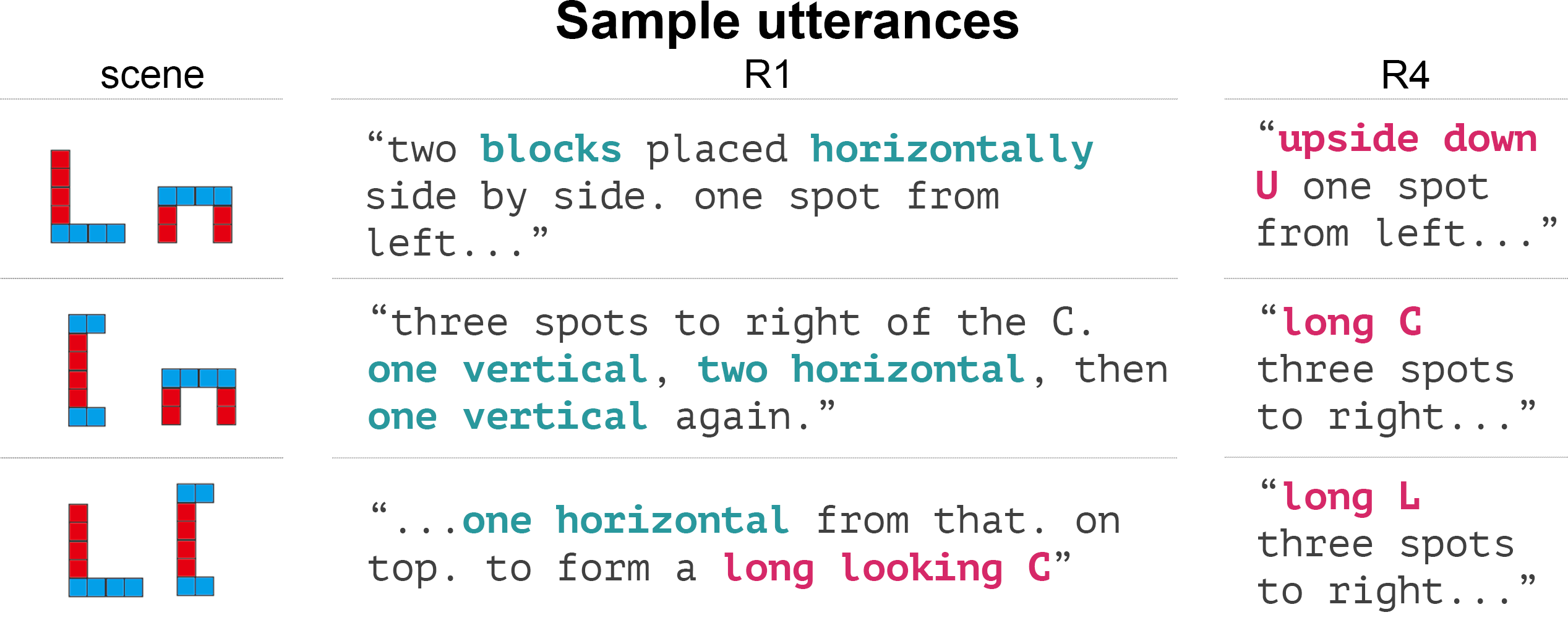}
    \caption[Emergence of tower-level expressions]{Example messages showing the emergence of tower-level expressions: \textit{upside down U}, \textit{long C}, and \textit{long L}.}
    \Description{A three-by-three table with columns labeled Scene, R1, and R4. In the first row, the scene shows an L-shaped tower and an upside-down U-shaped tower. R1 describes block-by-block placement, while R4 uses the tower-level phrase upside down U one spot from the left. In the second row, the scene shows a long C and an upside-down U. R1 gives a detailed sequence of block placements, while R4 says long C three spots to the right. In the third row, the scene shows an L and a long C. R1 describes individual block placements to form a long C, while R4 uses the tower-level expression long L three spots to the right.}
    \label{fig:study1_shift}
\end{figure}

\subsubsection{Communicative efficiency improves across repetitions}
Having shown that Builders could reconstruct the towers from the instructions, we next examined a basic signature of increasing abstraction in language (\autoref{fig:study1_shift}).
Because the same towers recurred, \rev{\textbf{(H2)}} we hypothesized that Instructors would exploit these regularities and provide more concise instructions over time. 
We analyzed both the number of words produced by the Instructor in each trial and the number of messages sent.
A mixed-effects model with a fixed effect of repetition and maximal random effects for items and participants revealed that Instructors used significantly fewer words ($\beta=-8.53$, $t(36.9)=-9.58$, $p< .001$; \autoref{fig:study1_all_results}B) and sent fewer messages ($\beta=-18.1$, $t(24)=-7.11$, $p< .001$) across repetitions.

\subsubsection{Level of referential abstraction increases across repetitions}
What allowed dyads to perform better while also using fewer words? \rev{\textbf{(H3)}}
We hypothesized that the increase in communicative effectiveness \rev{reflects a gradual shift toward higher-level, more abstract instructions.}
We first conducted a qualitative analysis to explore this possibility. 
We tokenized all of the Instructor's messages into individual words and examined, across our entire dataset, which words changed the most in frequency from the beginning to the end of the experiment (\autoref{fig:study1_all_results}C).
We observed that the frequency of low-level nouns like ``block'' and block-level modifiers like ``horizontal'' or ``red'' decreased the most, while that of high-level nouns like ``L'' or ``C'' and adjectives like ``tall'' increased the most. 

\rev{To assess how strongly dyads converged on a common vocabulary for tower-level abstractions, we computed the Jensen–Shannon divergence (JSD) between word-frequency distributions of dyads per repetition. The mean JSD increased from R1 to R4 (0.080, 95\% CI = [0.041, 0.118], $p = .004$), indicating that dyads’ vocabularies became \emph{more} divergent over time, and they developed distinct linguistic mappings from words to scene components.}

\rev{We next conducted} a more systematic analysis of message content. Four annotators, unaware of the study design and hypotheses, tagged each referring expression for block-level vs. tower-level references, yielding high agreement (intraclass correlation ICC = $0.83$, 95\% CI $=[0.82,0.84]$). 
We fit a mixed-effects model with fixed effects of repetition, expression type (tower vs. block), and their interaction, as well as maximal random effects for dyad. 
The significant interaction ($b=0.53$, $t(47.5)=4.8$, $p< .001$; \autoref{fig:study1_all_results}D) indicated that block-level references decreased while tower-level references increased across repetitions. 
Mean block-level references dropped from 7.3 to 3.6, whereas tower-level references rose from 0.6 to 1.1. 
The shift was primarily driven by an increase in tower-level references and a decrease in block-level references, as well as messages containing a mixture of both (\autoref{fig:study1_all_results}E).
\section{Physical Multimodal Study}
    In the second study, we examined a physical assembly task and changes in multimodal instructions (i.e., gesture and speech) over repeated interactions. Findings from the unimodal online study informed the design, including the selection of the task hierarchy, the number and shape of towers, and the number of repetitions to converge on task \rev{success}. 

    To balance internal validity and ecological validity, we chose an experimental paradigm that tightly controlled participant communication rather than co-located, unconstrained settings. We used augmented reality (AR) to constrain communication channels while multimodal messages were sent between Instructors and Builders. Builders received embodied, spatially aligned 4D gestures with synchronous audio, and Instructors received images of built structures indicating the extent to which instructions were correctly interpreted. Factorizing study variables enabled analysis of how information was adjusted across modalities and how conventions emerged, from which we built a computational model for future convention-aware agents that \rev{have} access \rev{to} these multimodal signals (i.e., audio and hand tracking).

\subsection{Method}
\subsubsection{Participants}
    We recruited $40$ participants\footnote{We use P1-P20 to refer to the 20 pairs. See Appendix A.1.1 for demographics.} ($23$ women, $14$ men, $3$ non-binary; ages $18$-$44$, $M = 23.68$, $SD = 5.05$) for our IRB-approved study. 
    All self-reported color vision (corrected as needed), sufficient hearing and manual dexterity for the study tasks, and fluency in English. Each session lasted \textasciitilde1~hr, with \$25 compensation.

        
\subsubsection{Stimuli}
\begin{figure}[h!]
    \centering
    \includegraphics[width=\linewidth]{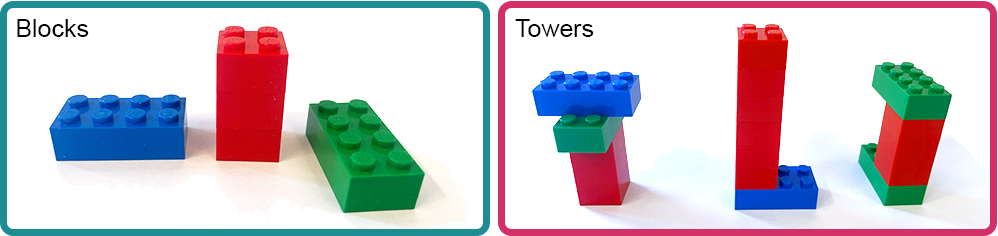}
    \caption[LEGO blocks and towers]{Physical LEGO blocks: blue (x-axis), red (y-axis), green (z-axis); and three-block towers: \textit{TREE} (unknown mapping, 3D), \textit{L} (alphabetic, 2D x-y), and \textit{C} (alphabetic, 2D z-y).}
    \Description{Photo of study materials. The top row labeled Towers shows three example three-block towers: TREE (red, green, then blue), L (blue, red, red), C (green, red, green). Bottom row labeled Blocks shows individual LEGO pieces: blue (sideways horizontal), red (vertical), and green (front and back horizontal).}
    \removegap
    \label{fig:legos}
\end{figure}

    The study included three block towers: \textit{C}, \textit{L}, and \textit{TREE}, each consisting of three physical LEGO blocks (\autoref{fig:legos}). In all towers, the orientation and size of each colored block were fixed: the blue block (2x4x1) aligned with the $x$-axis, red (2x2x4) with the $y$-axis, and green (2x4x1) with the $z$-axis. We chose LEGO building as an example of a spatial cognitive task~\cite{sheltonDCJHKL22}, with two simple 2D alphabetic towers (\textit{L} and \textit{C}) and one complex 3D tower (\textit{TREE}) that lacked an obvious linguistic \rev{convention} and used all three block primitives, making it difficult to represent with two hands.\looseness=-1

We chose to have four repetitions (R1–R4), based on the first study and prior work showing that conventions are typically formed within the first four repetitions~\cite{krauss1964changes}. \rev{Because scene-level abstractions were not formed or used in the unimodal study, we used only one tower per trial.} Each tower appeared once per repetition with a unique combination of position and orientation.\footnote{See Appendix A.1.2 for an example sequence.} The order of the twelve towers was randomized within and across repetitions.
    

\subsubsection{Apparatus}
    We examined how Instructors communicate to Builders using speech and gesture in a physically grounded task~\cite{kraut1982listener}, and how these modalities evolve. 
    Prior work shows that Builders' verbal \rev{(e.g., questions)} and \rev{implicit} non-verbal \rev{(e.g., pause)} feedback influences Instructor behavior~\cite{kraut1982listener}. Because this feedback can vary in amount and timing, we used an AR setup to tightly control the feedback from Builders \rev{and eliminate confounding variables}.\looseness=-1

\begin{figure}[h!]
    \centering
    \includegraphics[width=\linewidth]{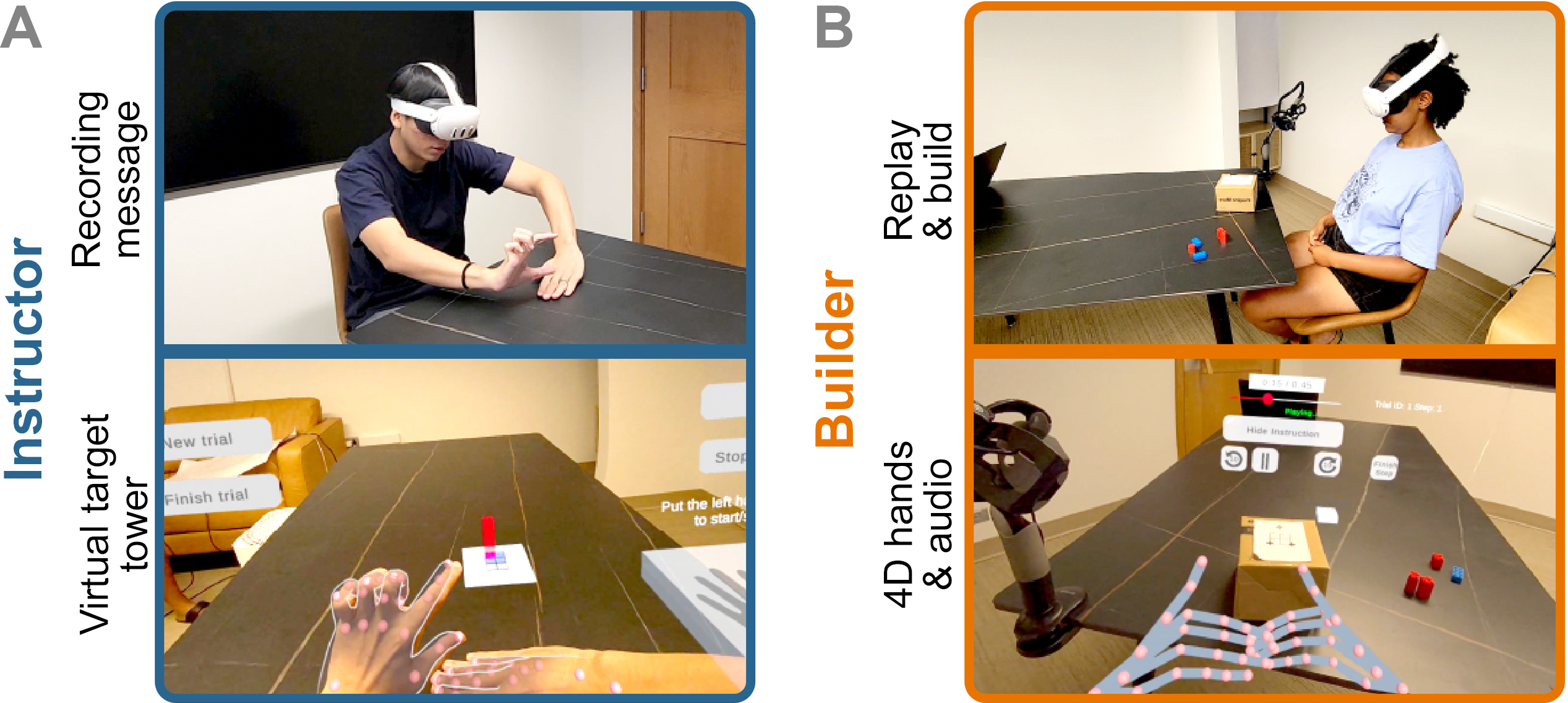}
    \caption[Study 2 system setup]{(A) The Instructor recorded a multimodal message (speech and gesture) in augmented reality (AR) describing how to build a virtual target tower with a specific pose. (B) The Builder replayed the message (audio and overlaid AR hands) and built the tower with physical blocks.}
    \Description{Two-panel image illustrating the setup. Panel A (Instructor): top shows a participant wearing a head-mounted display, seated at a table and making a c-shape gesture; bottom shows the Instructor’s AR view with semi-transparent tracked hands and a small virtual L-shape tower on a white 2 by 2 grid, plus on-screen controls (e.g., new/finish trial). Panel B (Builder): top shows a participant wearing the headset with blue/red physical LEGO blocks on the table; bottom shows the Builder’s AR view with a playback HUD (play/pause, progress bar) and articulated skeletal hands overlaid above the workspace while the physical blocks and grid are visible on the table. The images together depict recording a multimodal message (speech and gesture) and replaying it to guide physical assembly.}
    \removegap
    \label{fig:system}
\end{figure}

    Two participants in separate rooms wore Meta Quest 3 headsets while seeing their physical surroundings (Figure~\ref{fig:system}). The Instructor’s voice and hand movements were tracked, and a webcam captured images of the Builder’s environment after each step. The system transmitted audio and hand keypoints to the Builder and tower images to the Instructor. This setup isolated the Instructor’s speech and gestures, eliminated other non-verbal cues (e.g., eye gaze, facial expressions), and controlled the Builder’s feedback, enabling analysis of information transfer and modality use over time.


\begin{figure*}[t!]
    \centering
    \includegraphics[width=\textwidth]{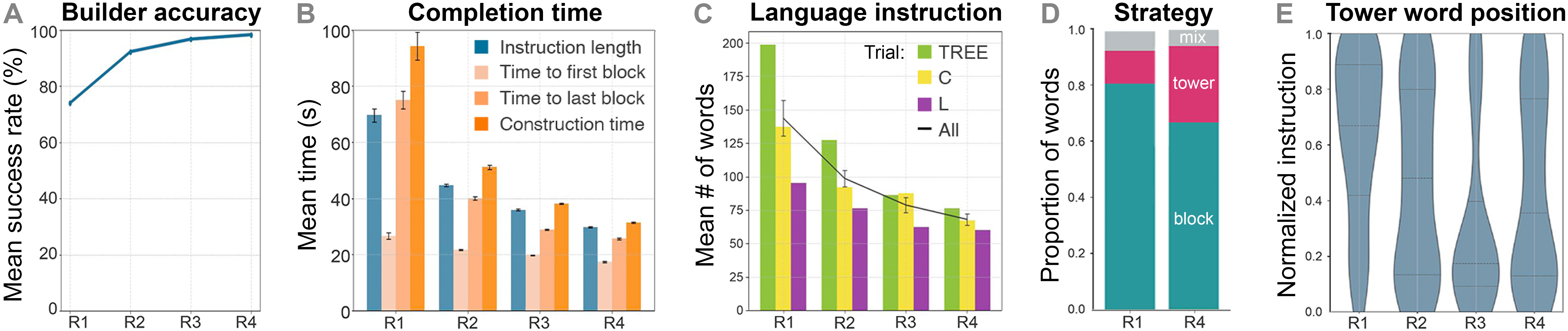}
    \caption[Study 2 results]{(A) Task performance in each repetition (R1-R4). (B) Instruction message length and builder time to place the first/last block, and complete the trial. (C) Number of words in instructions split by tower type. (D) Time proportion of block and tower words in R1 and R4. (E) Relative position of tower words during instructions. Error bars: standard error.}
    \Description{Results across repetitions R1-R4 in five panels. A) Line plot of mean builder success rate (percent) rising from roughly 70 to near the ceiling by R4. B) Shows grouped bars of mean time in seconds: instruction length and time to first and last block all decline across repetitions, with the largest drop for construction time. C) Shows bar charts for the mean number of words per trial by tower type (TREE, C, L) with an overlaid overall trend line; all decrease over repetitions. D) Shows stacked bars of the proportion of words labeled as block, tower, or mix; the block portion shrinks while the tower portion grows by R4. E) Shows violin plots of the normalized position of tower words within each instruction (0 start to 1 end). R1 starts narrow and then expands near the end, while R4 starts wide and shrinks near the end.}
    \removegap
    \label{fig:all_results}
\end{figure*}

    While prior work has explored gesture presentation methods, including 2D visualizations~\cite{benkoJW12, fussellKS00}, we presented 3D hand avatars with $24$ joint keypoints and bones to provide high-resolution information about hand position and depth. We displayed the hands from an egocentric perspective because pilot testing showed that third-person views introduced perspective-taking ambiguity (e.g., ``my left or your left?''), consistent with prior work~\cite{wang2021sa, jackson2006neural, koning2023investigating}. This choice also supports anticipated applications (e.g., AI-powered glasses) that infer intent from egocentric inputs and present instructions from this perspective~\cite{amoresBM15a, leeTKB18, youngLCMR19}. We used the \href{https://github.com/oculus-samples/Unity-DepthAPI}{Unity Depth API} to align virtual hands with the physical environment.


    For Builder feedback, pilot testing confirmed that 2D images were sufficient for this task. However, more complex towers or physical tasks may require live, photorealistic 3D renderings~\cite{johnsonSBXHSW23, kumaravelAFHG19, irlittiLZRHVV23, sakashitaKMW24}.

            

\subsubsection{Procedure}
    Participants were randomly assigned to the role of Instructor or Builder and worked in pairs in separate rooms. At the start of each trial, the Instructor viewed a 3D virtual twin of the target tower in AR, hidden from the Builder, and recorded multimodal messages (up to $2$ minutes). Instructors could send multiple messages or all instructions at once. The Builder replayed each message using basic controls (i.e., play/stop, forward/backward, and seek bar), then assembled the tower with physical LEGO blocks on a $2 \times 2$ grid.
    Upon completing a step, a camera automatically captured a photo of the reconstruction, which appeared in the Instructor’s AR view. The Instructor could then send a new message with the next steps, including corrections, or end the trial if the reconstruction was complete. After each trial, the experimenter evaluated the final tower as correct or incorrect, and the result was displayed in AR for both participants. Participants completed 12 trials and were instructed to construct towers as quickly and accurately as possible.\looseness=-1

\subsubsection{Analysis}\label{sec:multimodal_analysis}
    All Instructors' egocentric multimodal messages, 2D images of Builders' assemblies, and the experimenter's evaluations were saved for analysis. The entire session was also recorded by two external cameras to capture the Builders' and Instructors' actions. The footage of the construction was annotated with timestamps for each step. For analyzing speech, we first transcribed the audio in the messages with word-level timestamps using Whisper (large-v2)~\cite{radfordKXBMS23} and verified them manually for accuracy ($11$ of $256$ sentences were corrected). Two authors annotated references in R1 and R4 for level of abstraction (block, tower), clarity (clear, ambiguous), and information (shape, position, orientation). For analyzing gestures, we built a custom Unity-based desktop application for visualizing hand keypoints and bones in 3D space from different views, synchronized with the corresponding audio and transcript.\footnote{\rev{The multimodal dataset (audio transcripts and 4D gestures) and the custom desktop viewing app} are publicly available at: \href{https://multimodal-conventions.github.io}{https://multimodal-convention.github.io}} Two authors manually annotated right- and left-hand gestures in R1 and R4 based on their level of abstraction (block, tower), type (static, dynamic), and information (shape, position, orientation). Given the complexity of the high-dimensional, time-series data, instead of computing inter-rater reliability, we took a collaborative coding approach, where two authors discussed until they agreed on the coding or consulted a third author to resolve disagreements.\looseness=-1

\section{Multimodal Study Results}
\subsection{Task Performance}

\subsubsection{Success rate}
    Task success rate was defined as the proportion of valid messages in which Builders correctly constructed the towers. \rev{\textbf{(H1)}} We hypothesized that task success would improve across repetitions. The success rate improved from $74.03 \%$ in R1 to $98.36 \%$ in R4 with only one unsuccessful trial (Figure \ref{fig:all_results}A).




\subsubsection{Completion time}
    \textit{Instruction length} was the duration of the multimodal message that Instructors recorded. \textit{Construction time} was measured from when the Builders received an instruction to when they pressed the ``Finished'' button. The results are shown in Figure~\ref{fig:all_results}B, along with Builders' \textit{Time to first block} (from receiving an instruction to placing the first block) and \textit{Time to last block} (from receiving the instruction to placing the last block) in each trial. \rev{\textbf{(H2)}} We hypothesized that Instructors and Builders would become more efficient across repetitions. To test this, we used a linear mixed-effects model with one within-subject predictor (repetition) as a fixed effect and individual pairs as a random effect. There were significant effects of repetition on both instruction length ($\beta = -12.80, SE = 1.45, p < .001$), time to first block ($\beta = -2.99, SE = 1.10, p = .007$), time to last block ($\beta = -15.96, SE = 1.99, p < .001$), and construction time ($\beta = -20.19, SE = 2.73, p < .001$), indicating that instructions became shorter, and construction became faster.



\subsection{Convention Formation}

\subsubsection{Linguistic abstractions}
    We analyzed changes in the total number of words per trial and found that Instructors used fewer words to provide more concise instructions over time (Figure~\ref{fig:all_results}C). 
    We used a linear mixed-effects model with two within-subject predictors (repetition and tower shape) as fixed effects and individual pairs as a random effect to evaluate their impact on word count. The effect of repetition was significant ($\beta = -24.64, SE = 2.81, p < .001$). Moreover, instructions for \textit{L} had significantly fewer words ($\beta = -22.70, SE = 7.70, p = .003$), while instructions for \textit{TREE} had significantly more words ($\beta = 26.05, SE = 7.70, p = .001$). Although this gap narrowed across repetitions.

    Based on the unimodal study results, \rev{\textbf{(H3)}} we hypothesized that one explanation for this gain in linguistic efficiency is Instructors' utilization of abstractions that describe the entire tower shape with few words, which we refer to as ``tower words,'' chunking a sequence of block-based instructions. We found evidence for the use of tower words across repetitions, including ``C'' (82 times) and ``L'' (29 times). The \textit{TREE} tower was less commonly referred to as ``Tree'' (4 times), and more frequently as ``T'' (36) or ``Cross'' (21), likely due to the ambiguity of the mapping.\footnote{See Appendix~A.1.3 for the list of all tower words.}  
    Instructors establish conventions during the very first repetition, with $86.7\%$ of Instructors using at least one tower word in R1. To account for the reduction in message length from R1 to R4, we calculated the proportion of time spent describing block instructions and tower instructions (Figure~\ref{fig:all_results}D). We found that the proportion of tower instructions increased from $11.93\%$ in R1 to $27.29\%$ in R4. Using a permutation test ($10{,}000$ data shuffles) based on Euclidean distance (3D vector: block, tower, and other) for evaluating the difference between time proportions, we found this increase in proportion to be significant ($p = .015$).\looseness=-1
    
    We calculated the relative position of tower words ($w$) in messages ($M$) across all repetitions by dividing the index of $w$ by the number of words in $M$ (Figure~\ref{fig:all_results}E). We found that in R1, Instructors began by providing block instructions and then introduced tower words near the end of the trial. In later repetitions, they utilized these abstract tower words early in the trial. We used a linear mixed-effects model with one within-subject predictor (repetition) as a fixed effect and individual pairs as a random effect. We also included an orthogonalized quadratic effect of repetition to consider non-linearities. There were significant effects of repetition ($\beta = -0.44, SE = 0.096, p < .001$) and the orthogonalized quadratic term of repetition ($\beta = 0.071, SE = 0.019, p < .001$) on the relative positions of tower words. The results suggest that the position of tower words decreased, while the rate slowed across repetitions.\looseness=-1

    As Instructors shifted their strategy to abstract-first instructions, we also observed changes in Builders' construction strategy. In R1, most Builders ($19$ out of $20$) placed blocks one by one on the grid, while in R4, $11$ Builders constructed the entire tower, then placed it on the grid, reducing the gap between \textit{time to last block} and \textit{time to first block} in Figure~\ref{fig:all_results}B. This indicates that Builders also chunked their action execution, assembling towers during construction.\looseness=-1
    


\subsubsection{Gestural abstractions}
\begin{figure}[h!]
        \centering
        \includegraphics[width=\linewidth]{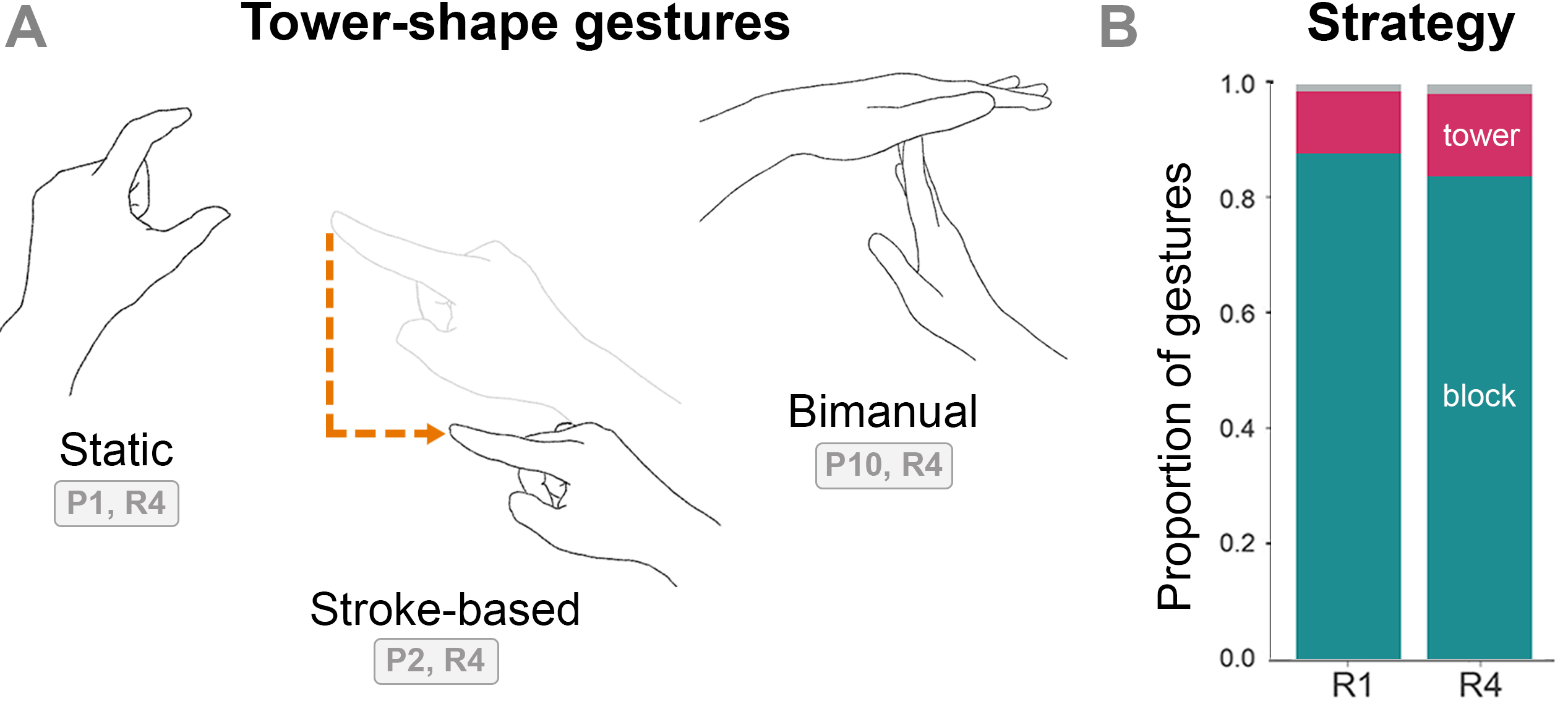}
        \caption[Gesture types]{(A) Tower-shape gestures were static and stroke-based, with bimanual gestures across both. (B) Time proportion of block and tower gestures in first and last repetitions.}
        \Description{A) Line drawings of tower-shape gestures: a static single-hand pose showing a C shape with index finger and thumb; a stroke-based gesture tracing an L path (dashed trajectory); and a bimanual gesture showing T with two open palm hands, each with example labels (P1, R4; P2, R4; P10, R4). B) Stacked bars for R1 and R4 showing gesture proportions: block around 0.9 in both, tower around 0.1 in both, indicating little change.}
        \removegap
        \label{fig:gestures}
    \end{figure}

    Gestures play an important role in multimodal communication. Similar to speech, we found evidence of block-level gestural instructions, with Instructors manipulating imaginary block pieces (see example in Figure~\ref{fig:teaser}), pointing to indicate a specific block position, or using iconic gestures to describe the entire tower shape. Overall, there were 78 instances of tower gestures (41 static and 37 stroke-based), 36 of which were bimanual (Figure~\ref{fig:gestures}A). We calculated the proportion of time participants spent performing block gestures and tower gestures in R1 and R4 (Figure~\ref{fig:gestures}B); however, we found no significant change in the proportions across repetitions. This suggests that participants used tower gestures in R1 to establish a tower-shape convention. In later repetitions, while they continued to use tower gestures, they increasingly referenced tower shapes with words alone.\looseness=-1

    \vspace{0.5cm}

\subsection{Multimodal Communication Dynamics}

\subsubsection{Informativeness of multimodal signals}
    We analyzed speech and gesture further to assess how much task-relevant information each modality encoded---what we refer to as the signals’ ``informativ\-eness''---and how this changed over repeated interactions. We focused on shape, position, and orientation, the spatial information necessary for the task. For each gesture and utterance, we coded whether they contained information about these parameters at the block or tower abstraction level. For those that contained information, we examined the interplay between the two modalities~\cite{de2006can}. We coded each multimodal reference as one of three categories: duplication of information across speech and gesture (redundant) to enhance comprehension~\cite{kelly2010}, use of hand gestures to disambiguate unclear utterances (complementary), such as ``like this'' or ``here''~\cite{iverson2005}, or speech without gestures (language-only).
    
\begin{figure}[h!]
    \centering
    \includegraphics[width=\linewidth]{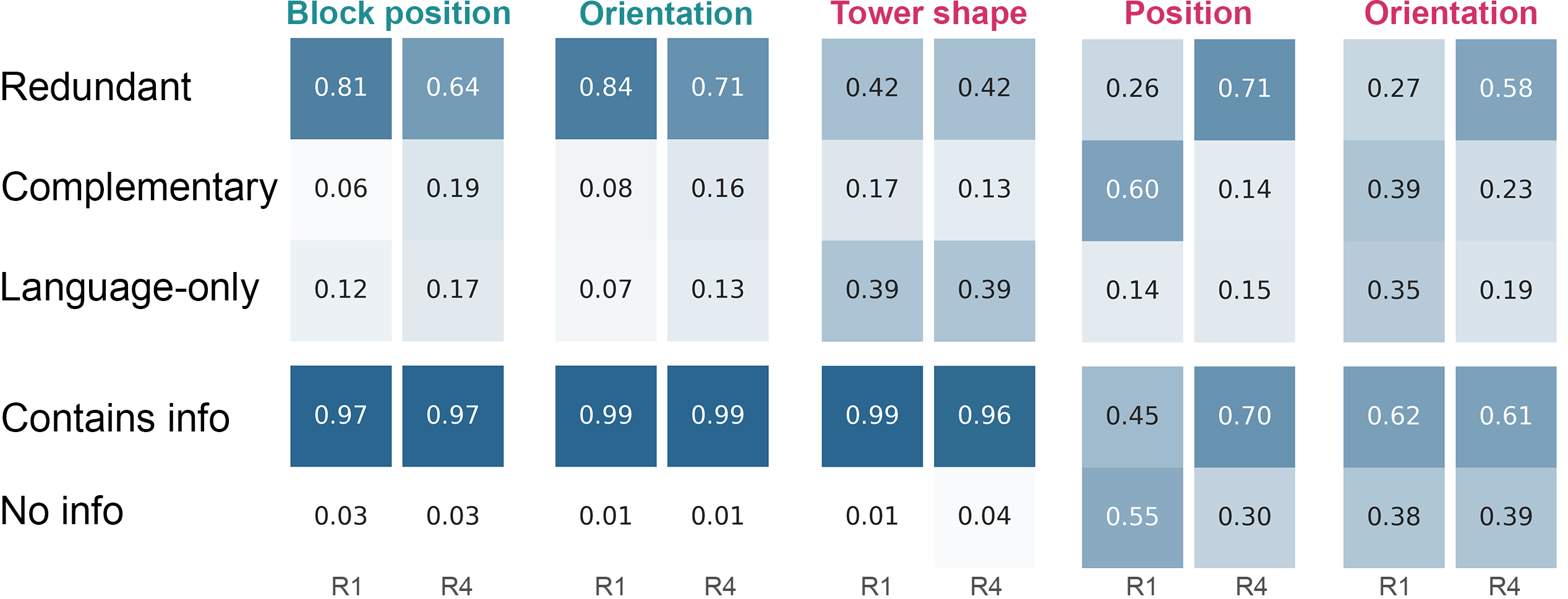}
    \caption[Instruction modality proportions]{Mean time proportion of redundant, complementary, and language-only instructions for position and orientation information of blocks and shape, position, and orientation of towers in the first and last repetitions.}
    \Description{Heatmap with 5 columns: Block position, Block orientation, Tower shape, Tower position, Tower orientation. Rows show Redundant, Complementary, Language-only, then Contains info and No info. Each cell lists the mean time proportions from R1 and R4. Block position: Redundant 0.81 to 0.64; Complementary 0.06 to 0.19; Language-only 0.12 to 0.17; Contains info 0.97 to 0.97; No info 0.03 to 0.03. Block orientation: Redundant 0.84 to 0.71; Complementary 0.08 to 0.16; Language-only 0.07 to 0.13; Contains info 0.99 to 0.99; No info 0.01 to 0.01. Tower shape: Redundant 0.42 to 0.42; Complementary 0.17 to 0.13; Language-only 0.39 to 0.39; Contains info 0.99 to 0.96; No info 0.01 to 0.04. Tower position: Redundant 0.26 to 0.71; Complementary 0.60 to 0.14; Language-only 0.14 to 0.15; Contains info 0.45 to 0.70; No info 0.55 to 0.30. Tower orientation: Redundant 0.27 to 0.58; Complementary 0.39 to 0.23; Language-only 0.35 to 0.19; Contains info 0.62 to 0.61; No info 0.38 to 0.39.}
    \removegap
    \label{fig:message_modality_type}
\end{figure}

    
    We calculated time proportions for each category (Figure~\ref{fig:message_modality_type}). For segments containing task information, we normalized the proportions so that \textit{redundant} + \textit{complementary} + \textit{language-only} = 1.0. \looseness=-1

\begin{figure*}[t!]
    \centering
    \includegraphics[width=\textwidth]{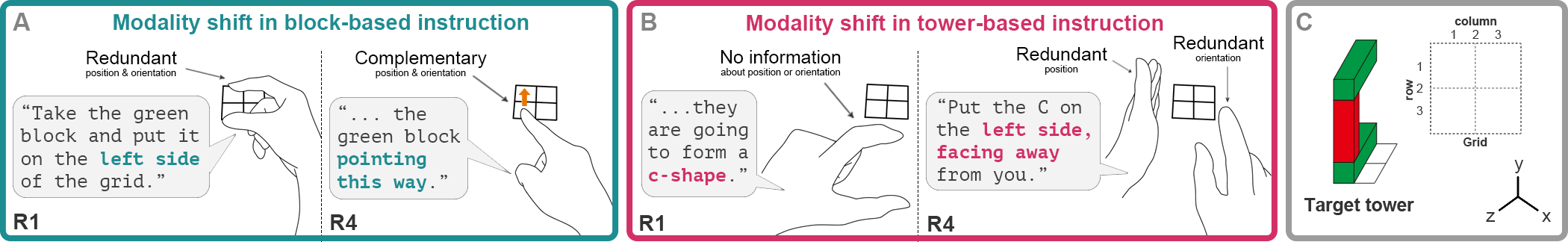}
    \caption[Modality shifts]{(A) Multimodal signals for position and orientation of blocks shift from redundant in R1 to complementary in R4. (B) For towers, no position or orientation information is provided when establishing an abstraction in R1, but redundancy is introduced in R4 to emphasize position and orientation changes. (C) Virtual target tower placed on the 2×2 grid.}
    \Description{Three-panel illustration of modality shifts. A) Modality shift in block-based instruction: R1 speech bubble says Take the green block and put it on the left side of the grid. A hand is holding an imaginary piece toward the left column of a 2×2 grid; the label reads Redundant, position and orientation. R4 speech bubble says the green block pointing this way. A hand is pointing near the bottom left cell of the grid with an arrow showing movement toward the top left cell; the label reads Complementary, position, and orientation. B) Modality shift in tower-based instruction: R1 speech bubble says they are going to form a C-shape. A c-shape hand pose with the index and thumb is shown far from the grid; the label reads No information about position or orientation. R4 speech bubble says Put the C on the left side, facing away from you. The right hand shows the C shape facing away, and the left hand with the palm open indicates placement on the left side of the grid; labels read Redundant, position, and Redundant, orientation. C) Target tower: a three-block green-and-red C-shape tower drawn beside a 2 by 2 grid with row and column markings (1,2, 3) and x (right), y (up), z (left) axes, depicting the reference frame used for position and orientation.}
    \label{fig:modality_shifts}
\end{figure*}

\subsubsection{Shifts in block-level instructions}
    We found that in R1, signals were predominantly redundant for both position ($81\%$) and orientation ($84\%$). However, redundancy decreased by $17\%$ for block position and $13\%$ for block orientation from R1 to R4. This shift indicates that, over time, Instructors favored concise deictic phrasing (e.g., ``this way'') with co-speech gestures for block-level instructions, rather than lengthy, precise utterances~\cite{bauerKS99} (Figure~\ref{fig:modality_shifts}A).


\subsubsection{Shifts in tower-level instructions}
    In R1, more than half of tower instructions lacked position information, and more than a third lacked orientation information. Instructors were likely establishing a shared convention for tower shapes after presenting detailed block-level instructions (Figure~\ref{fig:modality_shifts}B). In R1, only $26\%$ of tower position and $27\%$ of tower orientation information were redundant. However, redundancy increased by $45\%$ for tower position and $31\%$ for tower orientation. In R4, half of the Instructors used multimodal redundancy, providing more information than necessary~\cite{grice1975}, to emphasize variations from the previous repetition, which in our study was the randomized orientation and position of towers.
    To analyze these trends, we ran permutation tests ($10{,}000$ data shuffles). We found a significant shift in the distribution of time proportions for tower position from R1 to R4 ($p = .0009$). While time proportions for other categories (block position, block orientation, and tower orientation) also changed from R1 to R4, these shifts in distribution were not significant ($p = .085$, $p =  .20$, and $p = .16$, respectively).
    
\begin{figure}[h!]
        \centering
        \includegraphics[width=\linewidth]{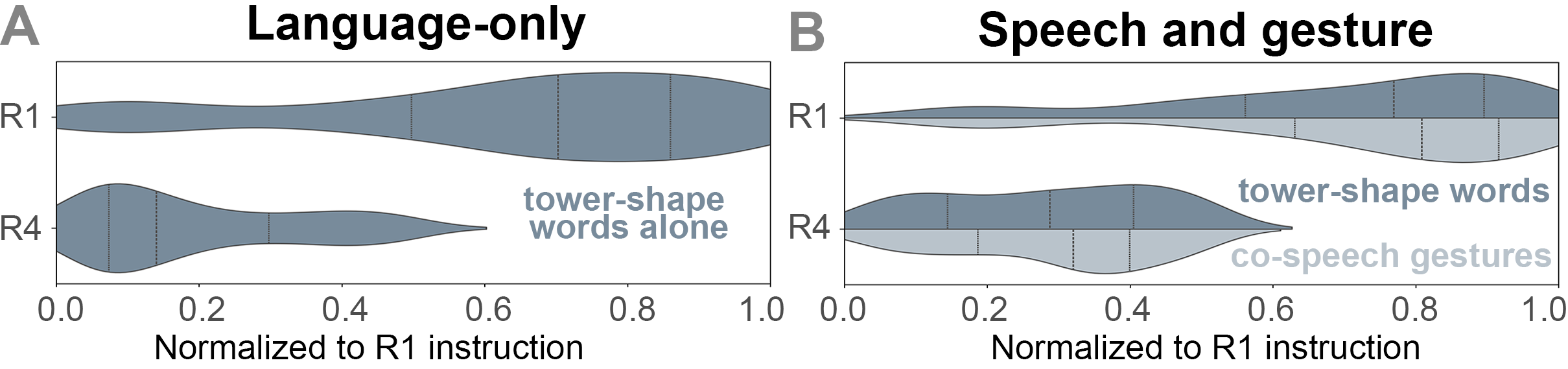}
        \caption[Position of tower words]{A: Relative position of tower words that appeared without gestures. B: Relative position of tower words and the corresponding redundant/complementary gestures.}
        \Description{Two sets of violin plots showing where tower-shape references occur within instructions (x-axis 0 = start, 1 = end; normalized to R1 length). Top, Language-only: R1 is narrow early and widens toward the end; R4 is wide near the beginning and tapers by mid-instruction. Bottom, Speech and gesture: in R1, co-speech gestures and tower-shape words spread across the instruction with greater width late; in R4, both shift earlier, but are spread more toward the center, and taper by mid instruction.}
        \removegap
        \label{fig:study2_violin_shift}
    \end{figure}

    When describing abstract tower shapes, gestures (redundant and complementary) typically followed tower words (Figure~\ref{fig:study2_violin_shift}). Co-speech gestures were used near the end of R1 to establish conventions, and in R4 were spread across shorter messages. Once conventions were established, in some trials, participants used tower words early on without relying on gestures. The cross-modal distribution of tower-shape information did not change across repetitions.
    
\subsubsection{User preferences and variability in modality shift}
\begin{figure}[h!]
        \centering
        \includegraphics[width=\linewidth]{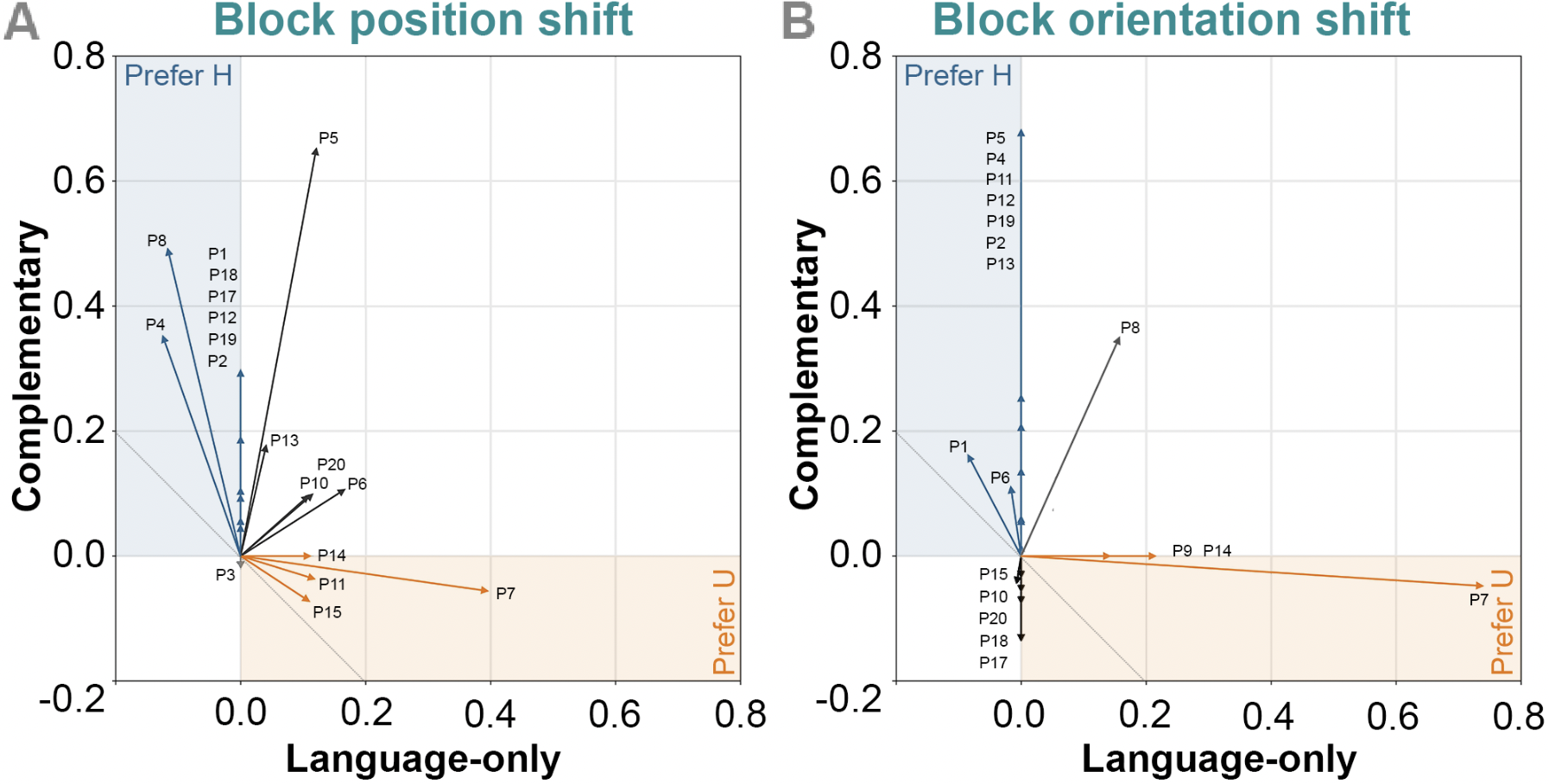}
        \caption[Block instruction shifts]{Shift in informativeness of block instructions from R1 to R4. Two participant groups diverged in their preferences: \textit{Prefer U} group shifted toward language-only instructions over time, and \textit{Prefer H} group shifted toward ambiguous language complemented with gestures.}
        \Description{Two plots showing participant-by-participant shifts from R1 to R4 in a 2D space with Language-only on the x-axis and Complementary on the y-axis. A) Block position shift: vectors start near the origin and either rise up and left toward the Prefer H region, or up and to the right in between, or right and down toward the Prefer U region. B) Block orientation shift: a similar pattern, with many moving upward, some moving right, and several staying near the origin but going slightly straight downward. The shaded quadrants indicate preferences for hand-gesture complementation (Prefer H) versus language-only instructions (Prefer U).}
        \removegap
        \label{fig:block_shifts}
    \end{figure}

    A more in-depth analysis of the shift in informativeness of block instructions revealed distinct participant groups that over time diverged in their modality preferences, consistent with prior work highlighting such individual differences~\cite{abramov2021relation}. For example, for block position, the first group, which we call the \textit{Prefer H} group, gravitated toward shorter, ambiguous utterances accompanied by hand gestures (Figure~\ref{fig:block_shifts}A). This \textit{Prefer H} group (P1, P2, P4, P8, P12, P17, P18, P19) was the largest subgroup, with $40\%$ of Instructors shifting from redundant to complementary use of multimodal signals. The smaller \textit{Prefer U} group (P7, P11, P14, P15) had the opposite preference and favored language-only instructions over time, eventually ceasing gesturing altogether. This reduced reliance on gestures may reflect gestural effort (e.g., arm fatigue) or growing familiarity (e.g., accumulated common ground on grid references and block orientation), making verbal instructions sufficient.

    We saw a similar trend for block orientation, with the \textit{Prefer H} group (P1, P2, P4, P5, P6, P11, P12, P19) being the largest subgroup at $40\%$. Unlike block position, however, $25\%$ of the dyads shifted toward more redundancy when describing the block orientation in the final repetition; moving downward in Figure~\ref{fig:block_shifts}B. We hypothesize that this difference reflects differential gesturing costs: orientation can be conveyed with in-place hand rotations, whereas position often requires reaching to the grid. This lower effort may explain the increase in redundancy observed for orientation, although this requires further investigation.

\section{Computational Model}

Toward developing convention-aware multimodal agents, \rev{we designed a unified computational model---a probabilistic (symbolic) system extending the Rational Speech Act (RSA) framework ~\cite{frank2012predicting,GoodmanFrank16_RSATiCS}---to explain behavior in both studies.}
To allow for complementarity between modalities, we adapted previous unimodal approaches~\cite{degen2020redundancy} to multimodal settings with both gesture and language. 
Our model allows agents to choose from redundant, complementary, or language-only messages and adjust their informativeness over time based on uncertainty about which communicative signals map to which abstractions. 
This enabled us to capture how participants adjusted their multimodal signals and explain the variability in how their communication preferences change over time.



\subsection{Domain-Specific Language}

We designed a domain-specific language (DSL) to describe agents' procedural knowledge about the assembly domain. 
This DSL $\mathcal{D}$ contains primitives for 3 colored blocks \dsl{R G B} and 9 positions \dsl{P}$_{\dsl{r}, \dsl{c}}$ across 3 rows and 3 columns on the grid. 
It also includes sub-towers (\dsl{T\_chunk}, \dsl{PL\_chunk}) and towers (\dsl{C\_chunk}, \dsl{L\_chunk},~\dsl{TR\_chunk}). 
For example, a C-shaped tower on the left can be expressed either as a sequence of primitives \dsl{G} \dsl{P}$_{\dsl{2}, \dsl{1}}$ \dsl{R} \dsl{P}$_{\dsl{3}, \dsl{1}}$ \dsl{G} \dsl{P}$_{\dsl{2}, \dsl{1}}$ or abstractly as \dsl{C\_chunk} \dsl{P}$_{\dsl{2}, \dsl{1}}$. \looseness=-1

\subsection{Multimodal Lexicon}

We assume each agent has a multimodal lexicon $l$ that maps utterances and gestures to symbols in the DSL, and a belief state ($\mathcal{L}$) over their collaborator's lexicon. We also assume that agents initially agree on the mappings of primitive block and position symbols. Paired agents must then resolve two sources of uncertainty over time: \rev{abstraction mapping ambiguity} and \rev{utterance ambiguity}. \looseness=-1
    
\subsubsection{Abstraction mapping ambiguity}
    Abstractions improve efficiency despite introducing uncertainty~\cite{fussellKS00}. To form ad hoc conventions, agents must coordinate on a mapping between utterances/gestures and the five chunked symbols at two levels (tower and sub-tower). This creates uncertainty over the $120$ possible lexicons ($|\mathcal{L}| = 5!$), each initialized with prior probability $\sfrac{1}{120}$.\looseness=-1

\subsubsection{Utterance ambiguity}
    The second source of uncertainty arises from agents' ability to combine modalities to produce complementary (ambiguous language + gestures) instructions. \rev{Using complementary gestures to specify position shortens instructions, despite adding linguistic ambiguity~\cite{fussellKS00}.} To account for ambiguities in language that arise from using gestures in a multimodal setting, we define two possible utterance states: \textit{clear} or \textit{ambiguous}. For example, \textit{``on the bottom half of the grid''} is \textit{clear} and \new{can be} decoded as \dsl{P}$_{\dsl{3}, \dsl{2}}$, while \textit{``here''} is \textit{ambiguous} and \new{requires gestures to} disambiguate the utterance. In this case, we assign a uniform probability of $\sfrac{1}{9}$ to each of the 9 possible positions on the grid.\looseness=-1
    

\subsection{Modality Preference in Instructions}
For each trial, the Instructor chooses a multimodal message to communicate their intention to the Builder. 
In the traditional RSA framework, a pragmatic speaker $S_1$ acts according to the speaker's utility function $U_s$, defined as:
    \begin{equation}
    U_s (u; t) = \log P_{L_0}(t|u) - C(u)
    \end{equation}
where $u$ is an utterance, $t$ is a target state, $L_0$ is a literal listener, who infers the target state $t$ given the literal meaning of $u$, and $C(u)$ is cost of $u$. 
The probability of generating $u$ is then proportional to $e^{U_s (u; t)}$\rev{, i.e., softmax selection over the message space}.\looseness=-1


We extend this utility function to consider more than one modality, specifically hand gestures in addition to language, and adapt to sequential signals. Given a target tower in trial $k$, the Instructor considers candidate sequential tower programs $(T_1^k, T_2^k, ...)$ acquired up to that point. Each $T_t^k$ consists of a set of steps $(t_{t, 1}^k, t_{t, 2}^k, ..., t_{t, |T_t^k|}^k)$ where $|T_t^k|$ is the number of steps ($s$) in the tower program $T_t^k$ and $t_{t, i}^k \in \mathcal{D}$. Then, the Instructor agent chooses a multimodal message $M_j^k = (m_{j, 1}^k, m_{j, 2}^k, ..., m_{j, |T^k_j|}^k)$ considering the literal Builder $B_0$ and the following utility function:



    \begin{equation}
    \label{equ:utility}
    \begin{split}
        U (M_j^k; T_j^k) = &\beta_i \sum_s \ln P_{B_0} (t_{j, s}^k | m_{j, s}^k) \\
        - &\beta_u (r) \sum_{s} C_u (u_{j, s}) - \beta_h (r) \sum_s C_h (h_{j, s})
    \end{split}
    \end{equation}

\noindent
where \new{$\beta_i$} controls how much the Instructor considers the informativeness relative to the cost. Each sub-message $m_{j, s}^k$ corresponds to a step $t_{j, s}^k$ and consists of an utterance $u_{j, s}$ and a gesture $h_{j, s}$. The assumption in RSA is that prior probability ($P_{B_0}$) is proportional to the literal meaning $l(m_{j, s}^k, t_{j, s}^k) = l(u_{j, s}, t_{j, s}^k) \cdot l(h_{j, s}, t_{j, s}^k)$.

While the original RSA formulation assumes that $l(m_{j, s}^k, t_{j, s}^k)$ is a binary value, we relax this assumption by introducing two continuous semantics $x_u, x_h \in [0, 1]$ as proposed by Degen et al.~\cite{degen2020redundancy}, allowing generation of redundant messages. 


    \begin{equation}
        l(u_{j, s}, t_{j, s}^k) = \begin{cases}
            x_u & \text{if $u$ is true of $t$} \\
            1 - x_u & \text{otherwise}
        \end{cases}
    \end{equation}
    
    \begin{equation}
        l(h_{j, s}, t_{j, s}^k) = \begin{cases}
            x_h & \text{if $h$ is true of $t$} \\
            1 - x_h & \text{otherwise}
        \end{cases}
    \end{equation}




    
In Equation~\ref{equ:utility}, the second and third terms on the right-hand side are cost functions where $\beta_u (r)$ and $\beta_h (r)$ are parameters that control the effect of utterances or hand gestures. These two parameters change depending on the repetition $r$ so that the model can simulate different combinations of utterances and gestures over time. High $\beta_u (r)$ leads to shorter, more ambiguous utterances, likely accompanied by gestures; and high $\beta_h (r)$ makes the agent prefer language-only instructions without gestures (i.e., $C_h (h_{j, s}) \rightarrow 0$).


After observing the message $M_j^k$, the pragmatic Builder $B_1$ assigns a symbol $t_{j, s}^k \in \mathcal{D}$ to $m_{j, s}^k$ with probability:

\begin{equation}
    \label{equ:b1_action}
    P_{B_1} (t_{j, s}^k | m_{j, s}^k) = \sum_{l \in \mathcal{L}} P_{I_1}(m_{j, s}^k | t_{j, s}^k, l) \cdot P_{I_1}(l)
\end{equation}



\noindent where $P_{I_1}$ refers to the Builder's belief about the pragmatic Instructor $I_1$ and is written with two terms using the utterance $u_{j, s}$ and the hand gesture $h_{j, s}$:

\begin{equation}
    \label{equ:multimodal_pi1}
    P_{I_1}(m_{j, s}^k | t_{j, s}^k, l) = \gamma \cdot P_{I_1}(u_{j, s} | t_{j, s}^k, l) + (1 - \gamma) \cdot P_{I_1}(h_{j, s} | t_{j, s}^k, l)
\end{equation}



\noindent where $\gamma$ determines how the model weighs utterances and gestures during inference. We assume that when a \textit{clear} utterance is used in a multimodal setting, the meanings of the utterance and hand gesture are aligned, producing the same probability distribution across the two terms above.


\subsection{Multimodal Abstractions}

If the Builder agent can infer the correct tower program, the Instructor agent may propose an abstract tower program $T^*$ with chunked symbols and send a new message $M^*$. Given that they have reached consensus, they update their prior beliefs following Bayes' rule: \looseness=-1

\begin{equation}
    \label{equ:b1_update}
    \log P_{I_1} (l | M^*) = \log P_{I_1} (l) + \sum_{m^* \in M^*} \log P_{I_1} (m^* | t^*, l) \qquad \forall l \in \mathcal{L}
\end{equation}

\begin{equation}
    \label{equ:i1_update}
    \log P_{B_1} (l | T^*) = \log P_{B_1} (l) + \sum_{t^* \in T^*} \log P_{B_1} (t^* | m^*, l) \qquad \forall l \in \mathcal{L}
\end{equation}

    
    

\noindent
where $t^*$ and $m^*$ are the corresponding symbol in $\mathcal{D}$ and sub-message, respectively. To correctly update beliefs over time, agents' $P(l)$ is normalized after each update, such that $\sum_{l \in \mathcal{L}} P(l) = 1$.

\subsection{Simulations}
    \subsubsection{Simulating abstractions} 
\begin{figure}[h!]
        \centering
        \includegraphics[width=0.8\linewidth]{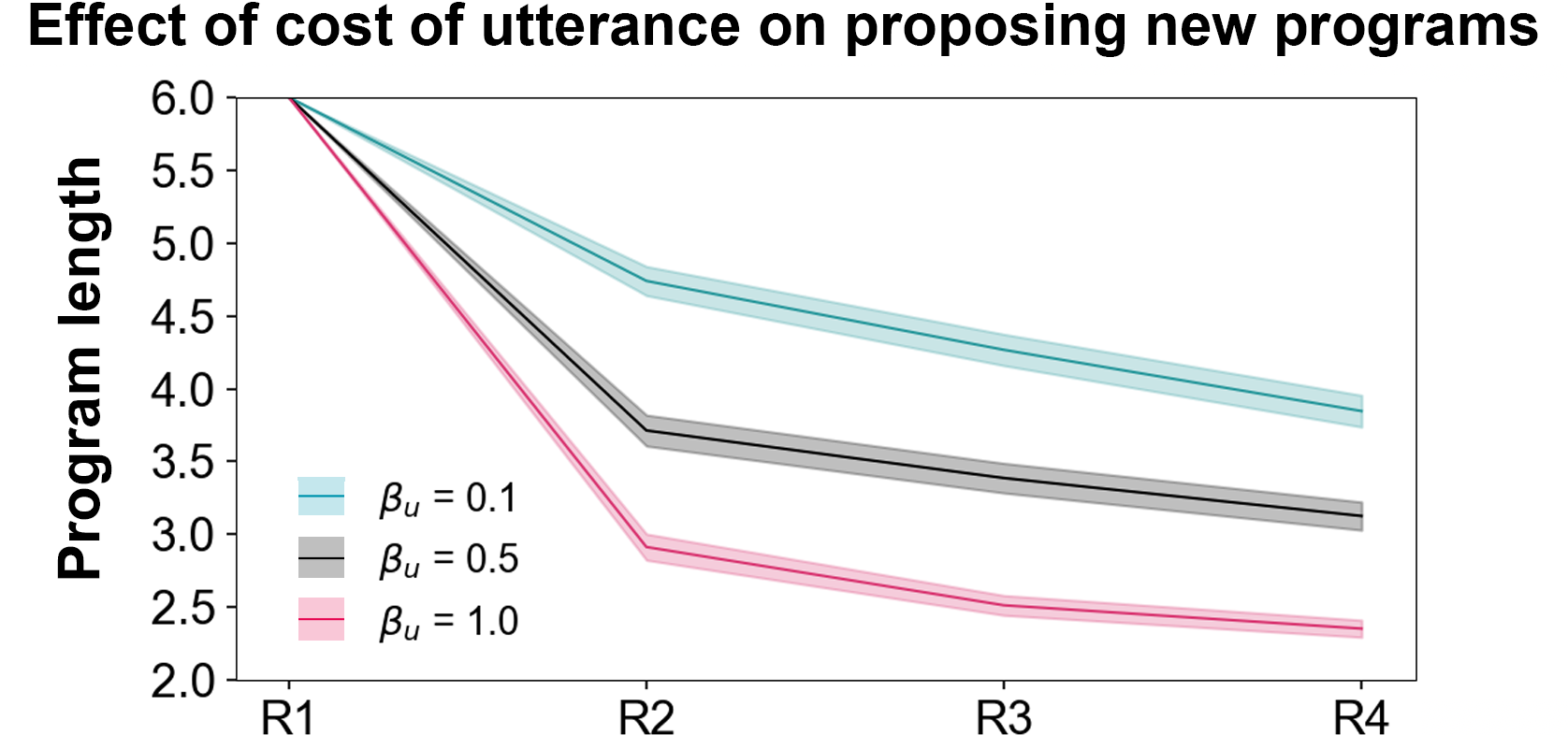}
        \caption[Simulation 1 results]{Simulation 1 results showing program length decreasing from R1 to R4 for $\beta_u$ values of 0.1, 0.5, and 1.0.}
        \Description{A) Line plot of program length across repetitions R1 to R4 for three utterance cost settings. High cost (βu = 1.0, pink) shortens programs rapidly from 6.0 at R1 to about 2.0 at R4. Medium cost (βu = 0.5, black) declines to about 3.0 by R4. Low cost (βu = 0.1, teal) changes little, staying near 4.0. Shaded bands show variability.}
        \removegap
        \label{fig:sim_1}
    \end{figure}

    In the first simulation, we asked whether the proposed model can acquire abstract tower representations over repetitions.
    For \rev{this}, we set $\beta_i = 0.3$ to prioritize minimizing cost over informativeness. To examine the effect of the cost of learning a new abstract program, we set $\beta_h = 0$ and varied $\beta_u$ to $0.1$, $0.5$, and $1.0$. The simulation setup was similar to the multimodal study, but tower positions and orientations were fixed across trials. The program length from $100$ simulations was recorded and aggregated for each repetition: a length of $6$ corresponds to no abstractions and $2$ to the agent using a tower program and a position symbol. The results, shown in Figure~\ref{fig:sim_1}, \rev{align with trends in both studies where program length decreased over time}.\looseness=-1 

\subsubsection{Simulating modality preferences}
\rev{We also asked whether the proposed model captures divergent modality preferences across participant groups.}
\rev{More specifically,} we examined whether the model could explain changes in Instructors' preferences among redundant, language-only, and complementary instructions over time. As an example, we focused on block position information. In the multimodal study, we observed two distinct participant groups: the \textit{Prefer U} group increasingly used language-only instructions from R1 to R4; the \textit{Prefer H} group increasingly used complementary instructions. \looseness=-1
    


Let $p_r^{obs}$, $p_u^{obs}$, and $p_c^{obs}$ denote the empirical probabilities of redundant, language-only, and complementary instructions $(p_r^{obs} + p_u^{obs} + p_c^{obs} = 1)$. In the model, the agent Instructor selects each type with probability $p_r^{pred}$, $p_u^{pred}$, and $p_c^{pred}$, proportional to $e^{U_r}$, $e^{U_u}$, and $e^{U_c}$, where utilities ($U_r$, $U_u$, and $U_c$) depend on parameter set $\theta = (\beta_i$, $\beta_u$, $\beta_h$, $x_u, x_h$). We fit $\theta$ to match the predicted distribution $(p_r^{\text{pred}}, p_u^{\text{pred}}, p_c^{\text{pred}})$  to the observed distribution for three conditions: R1, \textit{Prefer U} in R4, and \textit{Prefer H} in R4. This yields parameter sets $\theta_{r1}$, $\theta_{r4_u}$, and $\theta_{r4_h}$, which are then used for the simulation.

We fit the model parameters via Bayesian optimization to minimize the cross-entropy between predicted and observed distributions. 
The search ranges were: $\beta_u, \beta_h \in [0, 40]$ and $x_u, x_h \in [0.5, 1]$. 
We fixed $\beta_i$ at $10$, and held $x_u$ and $x_h$ constant at values found for $\theta_{r1}$ when estimating $\theta_{r4_u}$ and $\theta_{r4_h}$, so we could examine how $\beta_u$ and $\beta_h$ contribute to the agent's behavioral change compared to $\beta_i$. 
We performed $200$ optimization iterations with $40$ random initializations. \looseness=-1

\begin{figure}[h!]
        \centering
        \includegraphics[width=\linewidth]{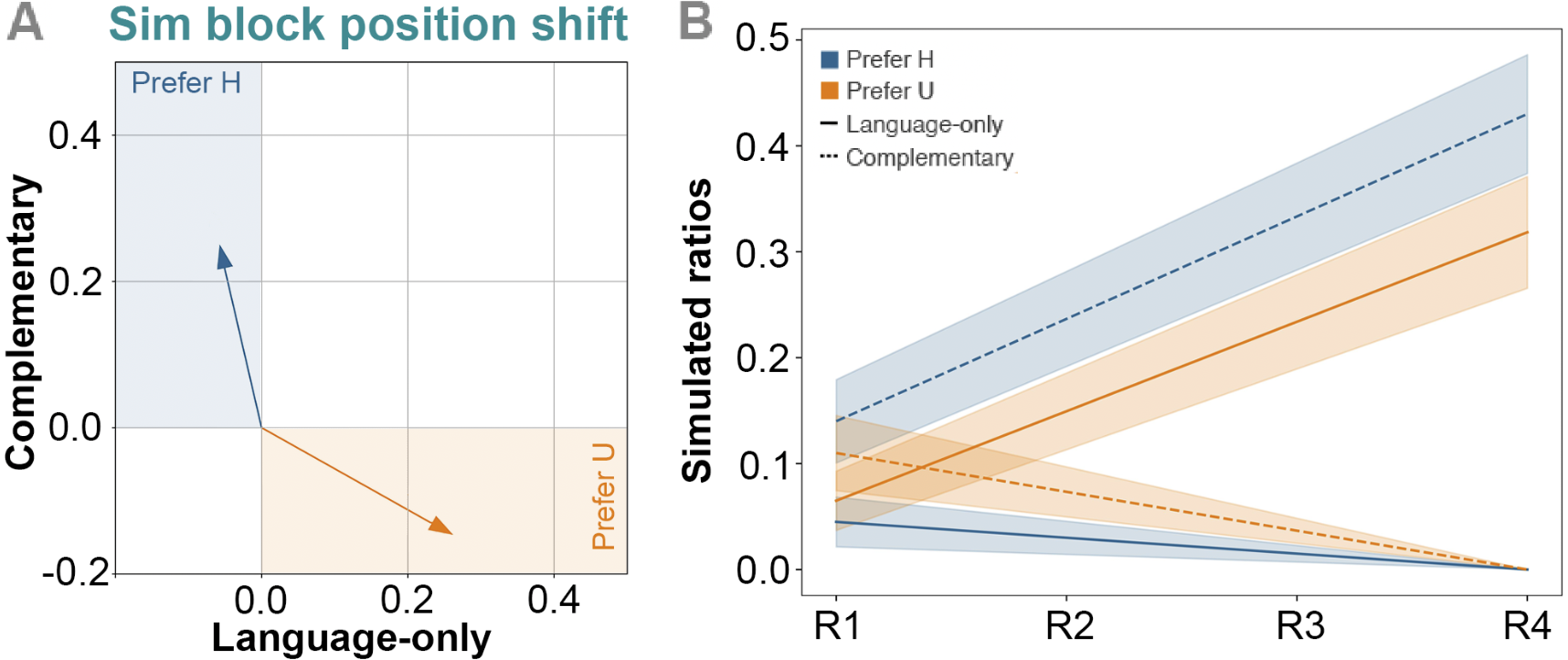}
        \caption[Simulation 2 results]{Simulation 2 results showing changes in language-only and complementary messages across repetitions.}
        \Description{A) Simulated block position shift. The x-axis is language-only, and the y-axis is complementary. Two vectors from near the origin: blue moves upward and slightly left into the shaded Prefer H region; orange moves right and slightly downward into the shaded Prefer U region. B) Simulated ratios across repetitions R1–R4 for two groups. Prefer H complementary (dashed blue) increases along with Prefer U language-only (solid orange). Prefer H language-only (solid blue) decreases along with Prefer U (dashed orange). Shaded bands show variability.}
        \removegap
        \label{fig:sim_new}
    \end{figure}

\rev{Bayesian optimization yielded fitted parameters for the R1: $\beta_u = 20.25$, $\beta_h = 9.23$, $x_u = 0.87$, and $x_h = 0.62$.
We held $x_u$ and $x_h$ constant when fitting R4 parameters.
For the \textit{Prefer U} group in R4, we obtained $\beta_u = 6.17$ and $\beta_h = 10.15$; for the \textit{Prefer H} group, $\beta_u = 21.01$ and $\beta_h = 3.09$.
These parameter shifts align with the observed behavioral changes. In \textit{Prefer U}, the decrease in $\beta_u$ and increase in $\beta_h$ reflect reduced concern for utterance cost and increased sensitivity to gesture cost, leading to language-only instructions.
Conversely, in \textit{Prefer H}, the decrease in $\beta_h$ reflects reduced sensitivity to gesture cost, leading to complementary instructions.  
We ran $200$ simulations for each parameter set and computed the mean proportions of complementary and language-only messages per repetition. 
Figure~\ref{fig:sim_new} shows how the probability of the agent choosing language-only or complementary instructions changes from R1 to R4 in the \textit{Prefer U} and \textit{Prefer H} conditions. 
The resulting probabilities closely match the observed data from Study 2.}

\section{Discussion}
    Effective collaboration in physical assembly tasks requires balancing communication cost and informativeness.
    In our iterated user study, we analyzed how Instructors’ multimodal signals changed across repetitions to convey intent. Pairs became faster and more accurate over time as Instructors reduced words per trial by introducing tower-level abstractions.
    We found that Instructors formed linguistic and gestural conventions near the end of the first exchange without specifying the position or orientation of abstract towers. In later trials, they named the tower words early to speed up construction time, and then added redundant speech and gesture to mark changes in position and orientation of the towers.\looseness=-1

    While people generally adopted task abstractions, \rev{conventions varied widely across dyads}. 
    At times, conventions were arbitrary (e.g., ``alligator'' or ``crocodile'' for \textit{C}; ``J'' for rotated \textit{L}), and even three-block towers were chunked differently (e.g., ``cross'' vs. ``T'' for \textit{TREE}). Modality strategies also diverged: some combined gestures with ambiguous language, while others minimized gesturing over time. To model these behaviors, we introduce a computational model that captures the formation of abstractions and balances informativeness with the costs of speech and gesture, and we show that it explains both the reduction in instruction length and participants’ diverging modality preferences. 
    
    
    Based on our findings, convention-aware agents for assembly tasks should learn users’ conventions for chunked instructions as they arise, over time default to abstract-first prompts when giving instructions, adapt modality to user preferences, and leverage redundancy to highlight changes from prior interactions. \looseness=-1 

\subsection{Limitations and Future Work}
    \subsubsection{Towards more complex physical structures}
        We focused on establishing an empirical foundation for multimodal convention formation using a limited set of block towers. However, people collaborate \rev{on} far more diverse \rev{and complex} physical structures in real-world settings. Future work should \rev{apply} our approach to \rev{richer assembly domains (e.g., furniture assembly) where parts vary more in shape and in how they can be combined.} This would reveal how well the principles identified here generalize to more \rev{complex tasks, such as those in EGGNOG~\cite{wangFNPM17} and HoloAssist~\cite{wangKRPCABFTFJP23}.}

    \subsubsection{From turn-based to real-time communication}
        We assigned consistent roles and controlled the turn-taking dynamics between dyads to draw precise links between the Instructor's communication and the Builder's building actions. However, in many real-world collaborative settings, the same individual might play multiple roles, sometimes giving instructions and other times following them. In addition, people collaborating are often able to communicate synchronously, which makes it possible for both communication and building actions to co-occur~\cite{zheng2024putting}, such as when a collaborator chooses to build before receiving a ``full'' set of instructions, or when someone interrupts their partner to ask a clarification question. Future work should investigate these more naturalistic settings where multimodal communication can be synchronous.\looseness=-1

        Prior work shows that immediate feedback enhances conversational grounding~\cite{gergleKF06, kraut1982listener}, whereas delayed feedback attenuates language adaptation~\cite{krauss1998language}. Accordingly, the changes we observed in our turn-based protocol may be less dynamic than in-person collaboration. Although our study was turn-based, our model is flexible and can be extended to synchronous settings by adopting finer-grained message units (e.g., word- or gesture-stroke level) and using contrastive inference to infer segment boundaries~\cite{degenKL21, kreissD20}. The framework can also incorporate lightweight, non-action feedback from Builders (e.g., uncertainty or confirmation signals) to influence instruction generation in real time. Evaluating these extensions is an important direction for future work.

    \subsubsection{Extensions of the computational model}
        Our model extends the RSA framework to generate and interpret abstract multimodal instructions and capture preferences in speech and gesture use. Several constants are currently manually set, but could be estimated online. For example, $\gamma$ can be derived from eye-tracking to determine whether Builders attend to gestures~\cite{gullberg2009attention, gullberg2006speakers}. Additionally, modality costs likely depend on task environment (e.g., grid size) and communication duration (e.g., arm fatigue), requiring further investigation.\looseness=-1
        
        We also made simplifying assumptions, including perfect semantic and temporal alignment between gestures and utterances~\cite{williamsO20} and a uniform prior over lexicons. Misalignment can make Instructors appear ``irrational,'' and cause pragmatic Builders to misinterpret intentions~\cite{kelly2010}; real-world agents would need to detect such misalignment using emerging conventions or accumulated common ground. Although some chunked instructions in our task are visually grounded (e.g., saying ``T'' while making a T-shaped gesture), we used a uniform prior to probe behavior in extreme cases.\looseness=-1
        

        Our model considers $120$ possible lexicons; scaling to more complex tasks will expand the set of possible abstractions and make belief updates more costly. Larger vocabularies and multimodal symbol spaces may require language-model approximations to pragmatic inference~\cite{degen2023rational, schusterAAD24}. The model also does not explain how people \emph{acquire} task abstractions or gestural conventions. Work on decomposing towers~\cite{mccarthy2023, sheltonDCJHKL22} and learning abstractions for complex artifacts~\cite{jonesGMR23, Koo_2022_CVPR} points toward ways to learn linguistic abstractions, but learning \emph{gestural} abstractions remains an open challenge.\looseness=-1

\section{Conclusion}
    We examined how conventions emerge in collaborative assembly across two settings. An online unimodal (text-only) study showed dyads shifting from block-by-block descriptions to concise tower-level abstractions, reducing instruction length over repetitions. A follow-up AR study in a physically grounded setting isolated speech and gesture: Instructors introduced tower-level abstractions near the end of the first exchange, then shifted to abstract-first instructions in later trials, using redundancy to mark changes in position and orientation. Pairs became faster and more accurate over time. To explain these behaviors, we introduced a computational model that captures abstraction formation and balances informativeness with the production costs of speech and gesture. Simulations reproduced shorter instructions and diverging modality preferences. Together, these results inform convention-aware agents for physical assembly tasks that learn users’ multimodal abstractions as they emerge, default to abstract-first instructions over time, add redundancy when conventions change or uncertainty is high, and personalize modality choices while adapting to changes in user preferences.\looseness=-1

\begin{acks}
    We thank the members of the Princeton HCI Group and the Cognitive Tools Lab at Stanford for fruitful discussions. Support for this work came from NSF CAREER \#2436199, an ONR Science of Autonomy Award, and a Hoffman-Yee Grant from the Stanford Center for Human-Centered AI awarded to J.E.F. Partial support was provided by Toyota RIKEN Fellowship to K.M.
\end{acks}
     
\bibliographystyle{ACM-Reference-Format}
\bibliography{References}

\appendix
\clearpage
\section{Appendix}

\subsection{Physical Multimodal Study}
    
\subsubsection{Participants}~\label{sec:appendix_demographics} Demographics information on Instructors and Builders in the multimodal study.


\begin{table}[h]
    \centering
    \caption[Participant demographics]{Participant demographics.}
    \Description{Table reporting participant demographics for Instructors and Builders. Rows list gender, age (mean and standard deviation), language fluency, handedness, AR/VR experience, and LEGO experience, with counts shown for each group.}
    \label{tab:study_participants}
    \begin{tabular}{p{1.8cm}p{2.8cm}p{2.8cm}}
         & Instructors & Builders \\ \midrule
        Gender & 6 M, 12 W, 2 NB & 8 M, 11 W, 1 NB \\
        Age & M $= 22.1$, SD $= 3.51$ & M $= 25.2$, SD $= 5.90$ \\
        Fluency & 20 Native & 9 Native, 11 Fluent \\
        Handedness & 18 Right, 2 Left & 18 Right, 2 Left \\
        AR/VR & 6 No experience & 8 No experience \\
        LEGO & 1 No experience & 1 No experience \\
    \end{tabular}
\end{table}

    
\subsubsection{Target Towers}
\begin{figure}[h]
    \centering
    \includegraphics[width=\columnwidth]{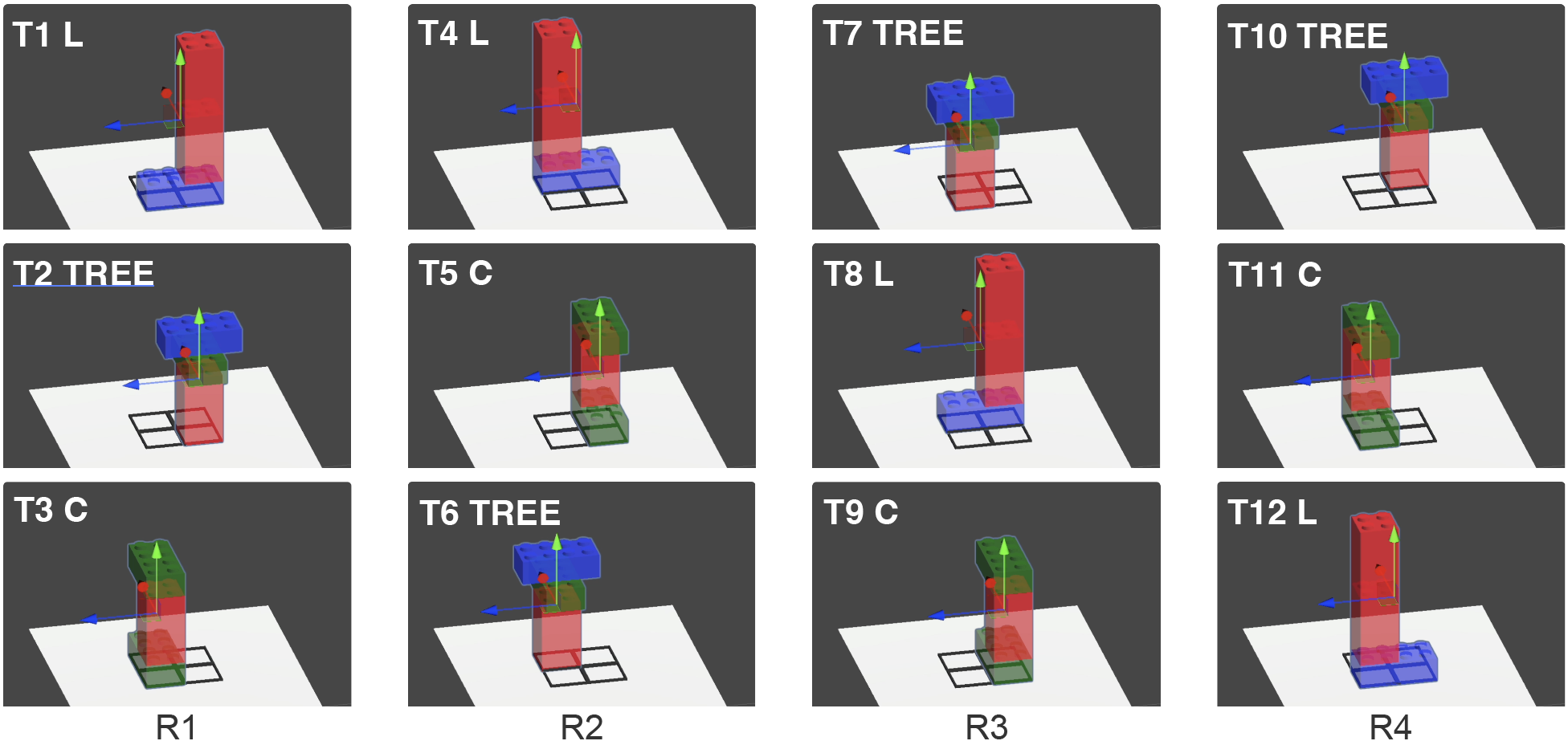}
    \caption[Target tower order]{An example order of the twelve target towers. Each tower type (\textit{C}, \textit{L}, \textit{TREE}) appears once in each repetition.}
    \Description{Twelve rendered target towers (T1–T12) shown on a table with a 2×2 grid and axis arrows, grouped by repetition. Each panel depicts a three-block tower with differing positions and orientations. R1: T1 L, T2 TREE, T3 C. R2: T4 L, T5 C, T6 TREE. R3: T7 TREE, T8 L, T9 C. R4: T10 TREE, T11 C, T12 L.}
    \label{fig:method_materials}
\end{figure}

    Figure~\ref{fig:method_materials} shows an example of the target towers. The study consisted of four repetitions (R1, R2, R3, R4), each with three trials (T), where the goal was to construct a tower. Each tower type (\textit{C}, \textit{L}, \textit{TREE}) appeared once in a repetition in a fully randomized order and with a different position and orientation.
    
\subsubsection{Analysis Tower Words}~\label{sec:appendix_tower_words}
    We extracted these words from the transcripts that refer to a tower/sub-tower shape.
    
    \begin{itemize}
        \item \textit{C}: c, c-shaped, c-shape, crocodile, alligator, alligators
        \item \textit{L}: l, l-shaped, l-shape, j
        \item \textit{TREE}: tree
        \begin{itemize}
            \renewcommand\labelitemii{$\circ$}
            \item \textit{T}: t, t-shaped, t-shape
            \item \textit{CROSS}: plus, cross, cross-like, cross-shaped
        \end{itemize}
        \item OTHER: structure, shape, construction
    \end{itemize}

\subsection{Computational Model and Simulations}

\subsubsection{DSL and Multimodal Signals}
    The following table shows the mapping between symbols in $\mathcal{D}$ and the corresponding multimodal signals. Each utterance has a cost $c_u$, which is proportional to the number of words. Hand gestures also have costs $c_h$, which are manually set. When the Instructor does not use gestures (utterance-only instructions), $c_h$ is set to $0$. Since we assume that the Instructor and Builder agents have a uniform prior, one of five random symbols (${\alpha, \beta, \gamma, \delta, \epsilon}$) is assigned to each chunked sub-tower/tower. The longer utterances for positions ($P_{i, j}$; cost: 0.6 or 0.7) refer to the statement \textit{"on the x of the grid"}, where $x$ is listed in the table below. 

\begin{table}[h]
    \centering
    \caption[Symbol–signal mapping]{Mapping between symbols and multimodal signals.}
    \Description{Table listing symbol primitives and their corresponding utterances and gestures, with associated costs for speech ($c_u$) and gesture ($c_h$).}
    \label{tab:dsl}
    \begin{tabular}{@{}p{0.7cm}p{4cm}p{3.2cm}@{}}
         & Utterance (cost, $c_u$) & Gesture (cost, $c_h$) \\ \cmidrule(lr){2-3}
        R & place a red block (0.4) & - (-) \\
        G & place a green block (0.4) & - (-) \\
        B & place a blue block (0.4) & - (-) \\
        $P_{1, 1}$ & top left (0.7), here (0.1) & point: top left (0.6) \\
        $P_{1, 2}$ & top half (0.7), here (0.1) & point: top half (0.6) \\
        $P_{1, 3}$ & top right (0.7), here (0.1) & point: top right (0.6) \\
        $P_{2, 1}$ & left half (0.7), here (0.1) & point: left half (0.6) \\
        $P_{2, 2}$ & middle (0.6), here (0.1) & point: middle (0.6) \\
        $P_{2, 3}$ & right half (0.7), here (0.1) & point: right half (0.6) \\
        $P_{3, 1}$ & bottom left (0.7), here (0.1) & point: bottom left (0.6) \\
        $P_{3, 2}$ & bottom half (0.7), here (0.1) & point: bottom half (0.6) \\
        $P_{3, 3}$ & bottom right (0.7), here (0.1) & point: bottom right (0.6) \\
        C & place a \{\textit{$\alpha,\beta,\gamma,\delta,\epsilon$}\} tower (0.4) & shape: \{\textit{$\alpha,\beta,\gamma,\delta,\epsilon$}\} (0.6) \\
        L & place a \{\textit{$\alpha,\beta,\gamma,\delta,\epsilon$}\} tower (0.4) & shape: \{\textit{$\alpha,\beta,\gamma,\delta,\epsilon$}\} (0.6) \\
        TR & place a \{\textit{$\alpha,\beta,\gamma,\delta,\epsilon$}\} tower (0.4) & shape: \{\textit{$\alpha,\beta,\gamma,\delta,\epsilon$}\} (0.6) \\
        T & place a \{\textit{$\alpha,\beta,\gamma,\delta,\epsilon$}\} tower (0.4) & shape: \{\textit{$\alpha,\beta,\gamma,\delta,\epsilon$}\} (0.6) \\
        PL & place a \{\textit{$\alpha,\beta,\gamma,\delta,\epsilon$}\} tower (0.4) & shape: \{\textit{$\alpha,\beta,\gamma,\delta,\epsilon$}\} (0.6) \\
    \end{tabular}
\end{table}

\subsubsection{Tower Programs}
\begin{figure}[h]
    \centering
    \includegraphics[width=\columnwidth]{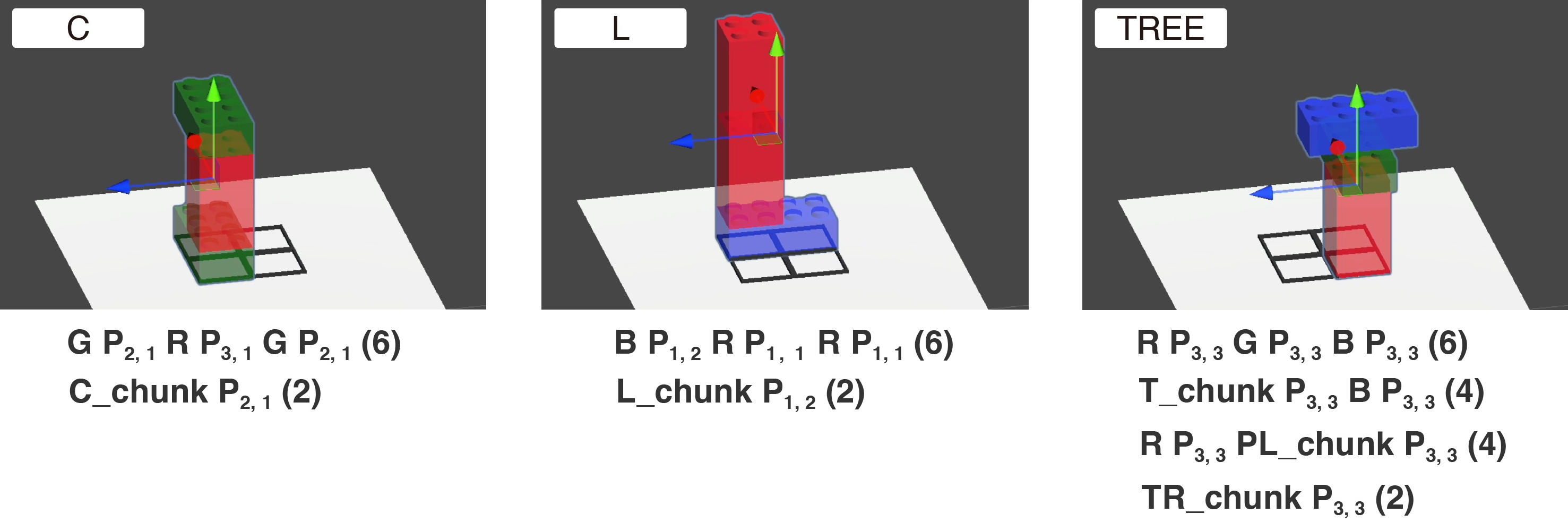}
    \caption[Tower programs]{Example target towers, tower programs, and their program length in parentheses, as used by the model.}
    \Description{Three rendered target towers with example instruction programs and chunks. Left, C: tower stands on left two cells; text below shows G P2,1 R P3,1 G P2,1 (6) and C_chunk P2,1 (2). Middle, L: tower stands on back row; text shows B P1,2 R P1,1 R P1,1 (6) and L_chunk P1,2 (2). Right, TREE: the red stem placed on the bottom right cell; text shows R P3,3 G P3,3 B P3,3 (6); T_chunk P3,3 B P3,3 (4); R P3,3 PL_chunk P3,3 (4); TR_chunk P3,3 (2).}
    \label{fig:appendix_tower_programs}
\end{figure}

    Figure~\ref{fig:appendix_tower_programs} shows target towers, tower programs, and their program length. In the simulation, the Instructor proposes a new tower program, from top to bottom, on each list.
 
\end{document}